\documentclass[seceq]{ptptex}

\notypesetlogo                       

\title{
Electron and nuclear pressures in electron-nucleus mixtures
}

\author{
Junzo \textsc{Chihara}$^{1,}$\footnote{E-mail: jrchihara@nifty.com} and Mitsuru \textsc{Yamagiwa}$^{2}$%
}

\inst{
$^1$ Higashi-ishikawa 1181-78, Hitachinaka, Ibaraki 312-0052, Japan\\
$^2$Advanced Photon Research Center, JAEA,
Kizu, Kyoto 619-0215, Japan
}

\abst{It is shown for an electron-nucleus mixture that the electron and nuclear 
pressures are defined clearly and simply by the virial theorem; 
the total pressure of this system is a sum of these two pressures. 
The electron pressure is different from the conventional 
electron pressure being expressed as the sum  of two times of kinetic energy and the potential energy in that 
the nuclear virial term is subtracted; this fact is exemplified 
by several kinds of definitions for the electron pressure enumerated 
in this work. The conventional definition of the electron pressure 
in terms of the nuclear virial term is shown inappropriate.
Similar remarks are made about the definition of 
the stress tensor in this mixture. It is also demonstrated 
that both of the electron and nuclear pressures become zero 
at the same time for a metal in the vacuum, 
in contrast to the conventional viewpoint that the zero pressure is realized 
by a result of the cancellation between the electron and nuclear pressures, 
each of which is not zero. On the basis of these facts, a simple equation of 
states for liquid metals is derived, and examined numerically 
for liquid alkaline metals by use of the quantum hypernetted chain equation 
and the Ashcroft model potential. 
}
\usepackage{epsf,namelist}
\begin{document} 

\maketitle

\section{\label{sec:int}Introduction} 
In order to determine the properties of solids, liquid metals and plasmas as 
electron-nucleus mixtures, 
the nuclei (ions) are usually treated as classical particles, which are considered as giving an external potential for the electrons: as a result, we can apply the density-functional (DF) theory to this inhomogeneous electrons. 
Thus, the DF theory provides the inhomogeneous properties of the electrons under this external potential. Therefore, some problems occur to determine 
homogeneous quantities of an electron-nucleus mixture including the nuclear 
contribution from the results of the DF theory, 
where the nuclei appear only as providing an external potential. 
Such an example of problems occurs in the determination of the total pressure of electron-nucleus mixtures on the basis of the DF theory. 
The electron pressure is frequently discussed from the DF theory; 
nevertheless, 
the term, `electron pressure' $\tilde P_{\rm e}$, is differently used 
in two ways in its 
meaning.
Some investigators \cite{Janak74,Averill81} take the electron pressure 
$\tilde P_{\rm e}$ for the electrons confined in a volume $V$ 
together with the fixed nuclei at the positions $\{{\bf R}_\alpha\}$
to be $3\tilde P_{\rm e}V=2T_{\rm e}+U_{\rm e}$ in terms of the exact kinetic and potential energies, 
$T_{\rm e}$ and $U_{\rm e}$, respectively. On the other hand, many kinds of definitions for 
the electron pressure lead to an expression 
$3\tilde P_{\rm e}V=2T_{\rm e}+U_{\rm e}-\sum_\alpha{\bf R}_\alpha\cdot{\bf F}_\alpha$ 
with use of the force ${\bf F}_\alpha$ on $\alpha$-nucleus, 
as will be shown in this work.
This difference is not only a problem of the definition, since 
the total pressure of an electron-nucleus mixture as a homogeneous system 
is to be determined from the electron pressure 
in conjunction with the nuclear contribution: the definition of electron pressure should be consistent with this purpose, as will be done in this work. 
Therefore, it is important to see the meaning of 
the electron pressure in the calculation of the total pressure taking account of the nuclear contribution. 

Following Slater,\cite{SlaterBK2} many investigators 
\cite{Janak74,Averill81,ZiescheGN88} have identified the electron 
pressure with the total force on the nuclei; 
$3\tilde P_{\rm e}V=\sum_\alpha{\bf R}_\alpha\cdot{\bf F}_\alpha$. 
As a consequence of this definition 
which leads to the relation $3\tilde P_{\rm e}V=2T_{\rm e}+U_{\rm e}$, the `electron pressure' 
becomes different from others as mentioned above.
Similarly, the same problem was seen in the definition of the stress tensors \cite{ZiescheGN88} in terms 
of the total force on the nuclei. 
This definition of $\tilde P_{\rm e}$ is based on the assumption: 
an electron-nucleus mixture confined in a volume  $V$ by wall potentials, $U_{\rm w}^{\rm ele}$ and $U_{\rm w}^{\rm nuc}$, for the electrons and the nuclei, respectively, is thermodynamically identical with the mixture with only the nuclei confined in $V$ by $U_{\rm w}^{\rm nuc}$, since in the latter system the electrons seem to spill out outside the nuclei a distance only on an atomic scale even if $U_{\rm w}^{\rm ele}=0$. In this work, we will examine the validity of this assumption  with use of the virial theorem which leads to a definition of the electron pressure in a more general form. From this new definition of the electron pressure, the total pressure of an electron-nucleus mixture is shown to be represented as a sum of the electron and nuclear pressures.

Also, we can give an important remark about the fact that the pressure in solids or liquid metals becomes zero in the vacuum;
 it is a standard viewpoint \cite{Price71,Hafner87} 
that the electron pressure is negative because of the electrons yielding a main part of 
cohesive energy, while the nuclear pressure is positive, and the total pressure becomes zero 
as a result of cancellation between two pressures in the vacuum. Contrary to this assertion, 
 the total pressure becomes zero, since  both the electron and nuclear 
pressures are zero at the same time in the vacuum, as we will see.
In this work, we discuss on these problems from fundamental and simple virial relations.

Based on the fact that the total pressure of an electron-nucleus mixture is a sum 
of the electron and nuclear pressures, we obtain a simpler pressure formula for 
a liquid metal, where the effective interaction among nuclei (ions) are 
approximately represented by a pair interaction; the nuclear pressure 
is expressed in the virial form using this pair interaction and the radial distribution 
function.
On the other hand, the electron pressure in a simple metal may be described by the jellium model. However, the electron pressure determined from the jellium model 
becomes zero only at the electron density of $r_s\approx 4$ and the jellium surface 
energy becomes negative for $r_s\approx 2$, while the jellium 
bulk modulus is negative at $r_s\approx 6$. These drawbacks of the jellium 
model are rectified by the stabilized jellium (SJ) model \cite{Perdew90} (or the ideal metal model \cite{Shore91,Shore99}), 
where an infinitesimally thin dipole layer is added to the surface of the 
uniform background in the jellium; this dipole layer produces a uniform field in 
the jellium to make it stable at any metallic densities. On use of this SJ model to 
obtain an electron pressure expression, 
we derived a pressure formula for a simple liquid metal in the present work. 
 
In the next section, we show that the electron and nuclear pressures are 
clearly defined on the basis of the simple and fundamental virial theorem, 
and construct three systems by changing the wall potentials confining 
the electrons and the nuclei in a finite volume. Based on these systems, 
we make several statements about the pressures for an electron-nucleus 
mixture. Similar remarks are made 
about the definition of the stress tensors for this mixture in \S\ref{sec:tensor}. In \S\ref{sec:def}, 
we enumerate several kinds of the definitions of the electron pressures 
proposed up to date. Equation of states (EOS) for liquid metals 
is set up in \S\ref{sec:deriv}; a numerical examination of this EOS is performed 
in \S\ref{sec:qhnc}. The last section is devoted to summary and discussion. In Appendix~\ref{append1}, the electron pressure formula is derived from the internal energy 
by performing a volume derivative in the framework of the DF theory.

\section{Virial theorem for an electron-nucleus mixture}\label{s:vir} 
The Hamiltonian for an electron-nucleus mixture is represented by
\begin{equation}\label{e:Hen}
\hat H=\sum_i\frac{\hat {\bf p}_i^2}{2m_i}
+\sum_\alpha\frac{\hat {\bf P}_\alpha^2}{2M_\alpha}
+\hat U(\{\hat{\bf r}_i\},\{\hat{\bf R}_\alpha\})
\equiv{\hat T_{\rm e}}+{\hat T_{\rm N}}+{\hat U}\,,
\end{equation}
where the coordinates and momenta of the electrons are denoted by $\{\hat{\bf r}_i,\hat{\bf p}_i\}$
and those of the nuclei by $\{\hat{\bf R}_\alpha,\hat{\bf P}_\alpha\}$
with the Coulomb interactions between particles
\begin{equation}
\hat U=\frac12\sum_{i\neq j}{e^2 \over |\hat{\bf r}_{i}-\hat{\bf r}_j|} 
-\sum_{i,\alpha}{Z e^2 
\over |\hat{\bf r}_{i}-\hat{\bf R}_{\alpha}|} 
+\frac12\sum_{\alpha\neq\beta}{Z^2 e^2 
\over |\hat{\bf R}_{\alpha}-\hat{\bf R}_{\beta}|} \,.
\end{equation}
This Coulomb potential $\hat U$ of an electron-nucleus mixture satisfies the following operator relation
\begin{equation}\label{e:Cpot}
\hat U=-\sum_i\hat{\bf r}_i\cdot\nabla_i\hat{U}
-\sum_\alpha\hat{\bf R}_\alpha\cdot\nabla_\alpha\hat{U}\,.
\end{equation}
Afterwards when the nuclei are regarded as classical particles, we use the symbols $\{{\bf R}_\alpha,{\bf P}_\alpha\}$ instead of $\{\hat{\bf R}_\alpha,\hat{\bf P}_\alpha\}$. 
In the standard approach to {\it homogeneous} liquid and solid metals, 
it is commonly assumed that the nuclei are treated as classical particles 
(or in the adiabatic approximation), and the electrons are considered as an {\it inhomogeneous} electron system under the external potential $U_{\rm ex}$ caused by the nuclei fixed at their equilibrium positions.
Here, it is important in treating an electron-nucleus mixture to recognize that there are four stages of approximations to the nuclear behaviour: (i) the nuclear mass is approximated as {\it infinite}, 
where the external potential $U_{\rm ex}$ is determined uniquely from a nuclear configuration, since the nuclei do not move but remain to stay at their equilibrium positions; (ii) the nuclei are approximated as classical particles in the sense that its position and momentum are commutable; (iii) the nuclei are regarded as quantum particles, but can be treated in the adiabatic approximation; and finally (iv) the nuclei can not be treated in the adiabatic approximation.
Therefore, we can apply the DF theory to the electrons in the electron-nucleus mixture without any approximation, if the nuclear mass is taken to be {\it infinite}.

Generally, the thermodynamic pressure $P$ of an inhomogeneous system caused by an external potential $U_{\rm ex}({\bf r})$ 
is clearly defined in terms of free energy $F$ or internal energy $E$, 
\begin{equation}\label{e:Pdev1} 
P=-\left.\frac{\partial F}{\partial V}\right|_{TNU_{\rm ex}({\bf r})}
=-\left.\frac{\partial E}{\partial V}\right|_{SNU_{\rm ex}({\bf r})}
\end{equation}
under the condition that the volume parameter $V$ of particles
 confined by a wall potential $U_{\rm w}({\bf r})$ 
 remains to specify the state as a 
natural variable when the external potential $U_{\rm ex}({\bf r})$ is imposed, as is described in detail.\cite{ChiharaUn} 
In addition to the thermodynamic pressure, there is another 'kinetic' definition of the pressure using a wall potential $U_{\rm w}$ on the basis of a standard assumption\cite{SlaterBK1,Mason69} concerning the relation between the force exerted by the wall ${U}_{\rm w}$ on the various particles and a hydrostatic pressure $P$:
\begin{equation}\label{e:gVir}
\sum_i\langle\hat{\bf r}_i\cdot\nabla_i{U}_{\rm w}\rangle=\oint_{\partial V}P{\bf r}\cdot d{\bf S}=P\oint_{\partial V}{\bf r}\cdot d{\bf S}=3PV.
\end{equation}
Furthermore, the thermodynamic pressure of this system is shown to be identical with the \lq kinetic' pressure in the form:
\begin{equation}\label{e:Gv}
3PV=-3V\left.\frac{\partial {F}}{\partial V}\right|_{TNU_{\rm ex}({\bf r})}
=2\langle\hat T\rangle-\sum_i\langle\hat{\bf r}_i\cdot\nabla_i\hat U_{\rm p}\rangle
\equiv \sum_i\langle\hat{\bf r}_i\cdot\nabla_i{U}_{\rm w}\rangle \,,
\end{equation} 
where the potential operator $\hat U_{\rm p}$ is interpreted as involving the external 
potential $U_{\rm ex}$, and $\hat T$ is the kinetic operator.\cite{Zwanzig50,Brown58,Mason69,Marc85}
Therefore, it is necessary from (\ref{e:gVir}) that the pressure takes a constant value at every point of the surface $\partial V$ 
for the volume parameter $V$ to be a natural variable. A simple example of this inhomogeneous system is an atom confined in a spherical wall with a nucleus fixed at its center.

[{\bf\,i-a\,}] As the first step to treat an electron-nucleus mixture on the basis of (\ref{e:Pdev1}) which afford to use the DF theory, let us consider a mixture of electrons and nuclei with the {\it infinite} nuclear mass, confined in a finite volume $V$ by wall potentials, $U_{\rm w}^{\rm ele}$ and $U_{\rm w}^{\rm nuc}$, respectively. Here, we assume this mixture constitutes a crystalline state in equilibrium, which is used as a model for the electron-band calculations. Since the nuclei remain at their fixed equilibrium positions, the nuclei make an external potential $U_{\rm ex}({\bf r})$ for the electrons, and the volume $V$ is a natural variable of this system. As a result, the 
pressure of this inhomogeneous electrons is determined by (\ref{e:Pdev1}) 
in terms of the electron free energy ${\cal F}_{\rm e}$, which is explicitly defined later in (\ref{e:Felec}). 
It is important to notice that the derivatives must be performed under the external potential being fixed.  Therefore, the nuclei producing the external potential $U_{\rm ex}$ for the electrons do not contribute to the pressure; 
the pressure is the electron pressure in the similar meaning to that of a noninteracting electron gas under the external potential. Since the electron free energy ${\cal F}_{\rm e}$  can be described exactly in terms of the DF theory,\cite{ChiharaUn} the electron pressure ${\tilde P_{\rm e}}$ is shown 
(see Appendix~\ref{append1}) to be 
\begin{eqnarray}
\hspace{-1.0cm}
{\tilde P_{\rm e}}&=&
-\left.\frac{\partial {\cal F_{\rm e}^{\rm DF}}}{\partial V}\right|_{TN_{\rm e}U_{\rm ex}({\bf r})}
=[2T_{\rm e}+U_{\rm e}
-\sum_\alpha{\bf R}_\alpha\cdot{\bf F}_{\alpha}]/3V  \label{e:ePdF1} \\
&=&[\, 2\langle \hat T_{\rm e}\rangle_{\rm e}-\sum_i\langle\hat{\bf r}_i\cdot\nabla_i\hat{U}\rangle_{\rm e}\, ]/3V=[\, \sum_i\langle\hat{\bf r}_i\cdot\nabla_i{U}_{\rm w}^{\rm ele}\rangle_{\rm e}\, ]/3V \,, \label{e:eP1}
\end{eqnarray}
where the bracket with a suffix e denotes the average over the electron states 
at a fixed nuclear configuration.
Really, this can be also proven from (\ref{e:Gv}) and (\ref{e:Cpot}).
Therefore, the electron pressure given above is a thermodynamic quantity, 
that is, a physical quantity defined {\it uniquely}.
Also, note that the electron pressure can be defined by a wall potential $U_{\rm w}^{\rm ele}$. However, it should be kept in mind that this pressure is 
not defined as a quantity of the homogeneous mixture.
 
[{\bf\,i-b\,}] In the second step to relate the above inhomogeneous system to a homogeneous 
electron-nucleus mixture, we consider this system to be a binary mixture of electrons and nuclei. 
Then, the total pressure of this homogeneous mixture is written as
\begin{eqnarray}
\hspace{-1cm} 3PV&=&2\langle \hat T_{\rm e}\rangle_{\rm e}-\sum_i\langle\hat{\bf r}_i\cdot\nabla_i\hat{U}\rangle_{\rm e}
-\sum_\alpha\langle{\bf R}_\alpha\cdot\nabla_\alpha\hat{U}\rangle_{\rm e} \nonumber\\
&\equiv&\sum_i\langle\hat{\bf r}_i\cdot\nabla_i{U}_{\rm w}^{\rm ele}\rangle_{\rm e}+\sum_\alpha{\bf R}_\alpha\cdot\nabla_\alpha{U}_{\rm w}^{\rm nuc} \nonumber\\
\hspace{-1cm}
&=&3{\tilde P}_{\rm e}V+\sum_\alpha {\bf R}_{\alpha}\!\cdot\!{\bf F}_\alpha=2T_{\rm e}+U_{\rm e} 
=\sum_i\langle\hat{\bf r}_i\cdot\nabla_i{U}_{\rm w}^{\rm ele}\rangle_{\rm e}-\sum_\alpha {\bf R}_{\alpha}\!\cdot\!{\bf F}_\alpha^{\rm w}  \,, \label{e:Om}
\end{eqnarray}
since the nuclei have infinite mass and do not move, but stay at their equilibrium positions. 
Note that the pressure of this homogeneous mixture is described in terms of the electron pressure ${\tilde P}_{\rm e}$ defined by (\ref{e:ePdF1}) for the inhomogeneous system. 
Also, all of the nuclear forces ${\bf F}_\alpha$ are zero except for nuclei near the surface in the short range of the wall potential, since the nuclei are in their equilibrium positions. On the other hand, the wall forces satisfying ${\bf F}_\alpha^{\rm w} =-{\bf F}_\alpha$ near the surface describe the nuclear pressure. Thus, we can see that the total pressure is composed of the electron and nuclear pressures.
In the derivation of (\ref{e:Om}), we have used the virial theorem for a homogeneous binary mixture confined in the volume $V$ by  
 a wall, which produces potentials, ${U}^{\rm A}_{\rm w}$ and ${U}^{\rm B}_{\rm w}$, for A and B particles, respectively; 
\begin{eqnarray}
\hspace{-1cm}3PV=-3V\left.\frac{\partial {F}}{\partial V}\right|_{TN}
\!\!&=&2\langle\hat T_{\rm A}\rangle-\!\sum_i\langle\hat{\bf r}_i\cdot\nabla_i\hat U_{\rm p}\rangle
+2\langle\hat T_{\rm B}\rangle-\!\sum_\alpha\langle\hat{\bf R}_\alpha\cdot\nabla_\alpha\hat U_{\rm p}\rangle  \nonumber\\
&\equiv&\sum_i\langle\hat{\bf r}_i\cdot\nabla_i{U}^{\rm A}_{\rm w}\rangle
+\sum_\alpha\langle\hat{\bf R}_\alpha\cdot\nabla_\alpha{U}^{\rm B}_{\rm w}\rangle \,.  \label{e:VirTh}
\end{eqnarray} 

[{\bf\,ii\,}] Furthermore, when the nucleus has a {\it finite} mass, but behaves as a {\it classical} particle which makes the relations in the adiabatic approximation exact, the electron pressure of an homogeneous electron-nucleus mixture is given by the configuration average over $\{{\bf R}_\alpha\}$ of the electron pressure in the inhomogeneous electrons under the external potential $U_{\rm ex}({\bf r}|\{{\bf R}_\alpha\})$:
\begin{eqnarray}\label{e:PdV}
3P_{\rm e}V=\left<-3V\left.\frac{\partial {\cal F}_{\rm e}}{\partial V}\right|_{TN_{\rm e}U_{\rm ex}({\bf r})}\right>_{\rm N}
=2\langle \hat T_{\rm e}\rangle-\sum_i\langle\hat{\bf r}_i
\cdot\nabla_i\hat{U}\rangle\,,
\end{eqnarray}
as will be discussed later in \S\ref{sec:dis}.
[{\bf\,iii\,}] Even when the nuclei behave as quantum particles, Eq.~(\ref{e:PdV}) is valid in the {\it adiabatic} approximation.\cite{ChiharaUn}
Therefore, this virial theorem (\ref{e:VirTh}) suggests that we can define the electron pressure in a general form even when the adiabatic approximation becomes invalid, as is described in the following.

[{\bf\,iv\,}] From the Hamiltonian (\ref{e:Hen}), we can obtain the following relation:
\begin{eqnarray}\label{e:Herm}
\langle\Psi_m|[\hat{\bf r}_i\cdot\hat{\bf p}_i,\hat H]|\Psi_m\rangle
=\imath\hbar\langle\Psi_m|\{\hat{p}_i^2/m-\hat{\bf r}_i\cdot\nabla_i\hat U\}
|\Psi_m\rangle=0\,,
\end{eqnarray}
due to the Hermitian property of the operator $\hat{\bf r}_i\cdot\hat{\bf p}_i$ of i-electron. 
Here, $\Psi_m$ denotes an eigenfunction of the Hamiltonian $\hat H$.
The ensemble average of the electron kinetic energy is related to the interacting potential 
by use of (\ref{e:Herm})
\begin{equation}\label{e:vE}
2\langle \hat{p}_i^2/2m\rangle=\langle\hat{\bf r}_i\cdot\nabla_i\hat{U}\rangle\,. 
\end{equation}
In addition, we obtain the similar relation for a nucleus in the system
\begin{equation}\label{e:vN}
2\langle \hat P_\alpha^2/2M\rangle=\langle\hat{\bf R}_\alpha\cdot
\nabla_\alpha\hat{U}\rangle\,.
\end{equation}
Here, the ensemble average of an operator $A$ is defined by
\begin{equation}\label{e:TreN}
\langle A\rangle \equiv {\rm Tr} [\exp(-\beta \hat H) A]/{\rm Tr} [\exp(-\beta \hat H)]\,,
\end{equation}
where the trace is taken over a complete set of 
states of the electron-nucleus mixture. 

Therefore, when the electrons and the nuclei are confined in a volume $V$ 
by {\it a wall} with wall potentials, $U_{\rm w}^{\rm ele}$ and $U_{\rm w}^{\rm nuc}$, respectively, 
Eqs.~(\ref{e:vE}) and (\ref{e:vN}) for this system are written 
in the following forms relating the electron and nuclear kinetic 
energies, $\hat T_{\rm e}$ and $\hat T_{\rm N}$, respectively: 
\begin{eqnarray}
2\langle \hat T_{\rm e}\rangle&=&\sum_i\langle\hat{\bf r}_i\cdot\nabla_i\hat{U}\rangle
+\sum_i\langle\hat{\bf r}_i\cdot\nabla_i{U}_{\rm w}^{\rm ele}\rangle\\ 
2\langle \hat T_{\rm N}\rangle&=&\sum_\alpha\langle\hat{\bf R}_\alpha\cdot\nabla_\alpha\hat{U}\rangle
+\sum_\alpha\langle\hat{\bf R}_\alpha\cdot\nabla_\alpha{U}_{\rm w}^{\rm nuc}\rangle\,,
\end{eqnarray}
where the wall potentials should be taken into account in the ensemble 
average (\ref{e:TreN}) by adding them to the Hamiltonian\cite{RiddelUlen50,Nishiyama51} or as the boundary condition\cite{Zwanzig50,Brown58,Mason69,Marc85} to solve the wave equation. At this point, it should be kept in mind that the interactions of the particles with the wall can be omitted from the new Hamiltonian if the boundary 
condition that $\Psi=0$ for any particle on the wall is made.\cite{Brown58,Mason69,Marc85}
As a result, with use of the Hamiltonian (\ref{e:Hen}) and this boundary condition, the electron and nuclear pressures are expressed respectively in the forms:
\begin{eqnarray} 
3P_{\rm e}V &\equiv&\sum_i\langle\hat{\bf r}_i\cdot\nabla_i{U}_{\rm w}^{\rm ele}\rangle
=-\sum_i\langle\hat{\bf r}_i\cdot{\bf F}_i^{\rm w}\rangle\label{e:Pe}\\
&=&2\langle \hat T_{\rm e}\rangle-\sum_i\langle\hat{\bf r}_i\cdot\nabla_i\hat{U}\rangle
=2\langle \hat T_{\rm e}\rangle+\langle\hat{U}\rangle-\sum_\alpha\langle\hat{\bf R}_\alpha\cdot\hat{\bf F}_\alpha\rangle\label{e:Pe2}\\
3P_{\rm N}V &\equiv&\sum_\alpha\langle\hat{\bf R}_\alpha\cdot\nabla_\alpha{U}_{\rm w}^{\rm nuc}\rangle
=-\sum_\alpha\langle\hat{\bf R}_\alpha\cdot{\bf F}_\alpha^{\rm w}\rangle\label{e:Pn}\\
&=&2\langle \hat T_{\rm N}\rangle-\sum_\alpha\langle\hat{\bf R}_\alpha\cdot\nabla_\alpha\hat{U}\rangle \label{e:Pn2}\,,
\end{eqnarray} 
since the external virials, $\sum_i\langle\hat{\bf r}_i\cdot\nabla_i{U}_{\rm w}^{\rm ele}\rangle$ 
and $\sum_\alpha\langle\hat{\bf R}_\alpha\cdot\nabla_\alpha{U}_{\rm w}^{\rm nuc}\rangle$, 
enable us to define the electron and nuclear pressures, respectively, on the base of the standard assumption: 
$\sum_i\langle\hat{\bf r}_i\cdot\nabla_i{U}_{\rm w}\rangle=\oint_{\partial V}P{\bf r}\cdot d{\bf S}$.\cite{SlaterBK1,Mason69,Marc85}. 
For the volume parameter $V$ to be a natural variable, 
the pressures, $P_{\rm e}$ and $P_{\rm N}$, must take a constant 
value at every point of the surface $\partial V$.
In the above, Eq.~(\ref{e:Pe2}) is derived by the relation (\ref{e:Cpot}), and the force on $\alpha$-nucleus is given by $\hat{\bf F}_\alpha=-\nabla_\alpha \hat U$. 
The definition of the electron pressure (\ref{e:Pe}) can be used even when the nuclei behave as quantum particles {\it without use of any approximation}. 

At this stage, we can construct the following three systems of electron-nucleus mixtures 
by choosing the electron and nuclear wall potentials in different ways.

(I) {System-I}, where the electrons and nuclei are confined in the volume $V$.
In this system, the total pressure is given by
\begin{eqnarray} 
\hspace{-1cm} 3PV&=&3(P_{\rm e}+P_{\rm N})V
=2\langle \hat T_{\rm e}\rangle+2\langle \hat T_{\rm N}\rangle+\langle\hat{U}\rangle\equiv 3P_{\rm I}V \,.\label{e:PI1} 
\end{eqnarray} 

(II) {System-II}, where $U_{\rm w}^{\rm ele}=0$ 
and $U_{\rm w}^{\rm nuc}\neq 0$. 
In this system, the electron pressure becomes zero because of (\ref{e:Pe}):
\begin{equation}\label{e:Pe0II}
3P_{\rm e}V_{\rm e}=0
=2\langle \hat T_{\rm e}\rangle+\langle\hat{U}\rangle-\sum_\alpha\langle\hat{\bf R}_\alpha\cdot\hat{\bf F}_\alpha\rangle\,.
\end{equation}
Therefore, the total pressure of this system can be attributed only to the nuclear part:
\begin{eqnarray}
3PV&=&2\langle \hat T_{\rm N}\rangle-\sum_\alpha\langle\hat{\bf R}_\alpha\cdot\nabla_\alpha\hat{U}\rangle
=2\langle \hat T_{\rm N}\rangle+\sum_\alpha\langle\hat{\bf R}_\alpha\cdot\hat{\bf F}_\alpha\rangle \label{e:PII1}\\ 
&=&2\langle \hat T_{\rm N}\rangle+2\langle \hat T_{\rm e}\rangle+\langle\hat{U}\rangle\equiv 3P_{\rm II}V\,,  \label{e:PII2}
\end{eqnarray}
At this point, it should be noted that the zero pressure relation (\ref{e:Pe0II}) is used in the derivation of (\ref{e:PII2}) from (\ref{e:PII1}).

(III) {System-III}, where $U_{\rm w}^{\rm nuc}= 0$ and $U_{\rm w}^{\rm ele}= 0$. 
This system represents nothing but an electron-nucleus mixture in the vacuum, where $P_{\rm e}=0$ {\it and} $P_{\rm N}=0$ due to (\ref{e:Pe}) and (\ref{e:Pn}). Consequently, the 
total pressure of this system becomes zero: $P=P_{\rm e}+P_{\rm N}=0$.\\

On the basis of the fundamental relations derived here in conjunction with System-I, II and III constructed in the above, we can make the following statements from (A) to (E):
\begin{namelist}{(X)}
\item[(A)]: The electron pressure $P_{\rm e}$ and the nuclear pressure $P_{\rm N}$ can be 
clearly defined for an electron-nucleus mixture by (\ref{e:Pe2}) 
and (\ref{e:Pn2}), respectively; the total pressure is a sum 
of the electron and nuclear pressures, as shown in (\ref{e:PI1}).
\item[(B)]: The electron pressure is described by (\ref{e:Pe2}), where the nuclear virial is subtracted.
\item[(C)]: The pressure of System-I 
is the thermodynamic pressure of an electron-nucleus 
mixture. It has been shown that the pressure defined by the sum of (\ref{e:Pe2}) and (\ref{e:Pn2}) becomes identical with the thermodynamic pressure determined by the volume derivative of free energy.\cite{RiddelUlen50,Nishiyama51,Zwanzig50,Brown58,Mason69,Marc85}\,  
\item[(D)]: The pressure $P_{\rm I}$ in System-I has an identical expression to $P_{\rm II}$ in System-II. 
Therefore, some investigators\cite{Janak74,Averill81,ZiescheGN88} believe that System-II is identical to 
System-I in the sense that both systems provide the same thermodynamic quantities (e.g., the pressure).
However, the pressures, $P_{\rm I}$ and $P_{\rm II}$, are equal with each other 
only when these systems are nearly equal to the system-III ($P_{\rm e}=P_{\rm N}=0$). 
It should be noticed that System-II is in general an inhomogeneous system, since this model involves a superatom in the extreme limit ($V\to 0$) with all nuclei compressed 
to become a point charge as this mathematical model. On the other hand, System-I is always homogeneous, since the wall effect can be replaced by periodic boundary conditions for this system.  
Therefore, System-I and System-II are
different from each other in general, although (\ref{e:PI1}) and (\ref{e:PII2}) have the identical expression; 
a {\it physical quantity} $P_{\rm e}$ is not zero in the one and zero, in the other as Eq.~(\ref{e:Pe0II}) indicates.
\item[(E)]: A metal in the vacuum is {\it defined} as an electron-nucleus mixture without the wall potentials to confine the electrons ($U_{\rm w}^{\rm ele}=0$) and the nuclei ($U_{\rm w}^{\rm nuc}=0$) in a finite volume.
Therefore, {\it by definition} the thermodynamic pressure of a metal in the vacuum is 
zero, which is realized by the condition $P_{\rm e}=0$ {\it and} $P_{\rm N}=0$, as System-III exhibits: this results from the virial theorem for {\it single} particle, (\ref{e:vE}) and (\ref{e:vN}).
This fact makes a contrast to the usual treatment of a metal where the pressure of a 
metal becomes zero as a result of the cancellation in a sum: $P=P_{\rm e}+P_{\rm N}=0$ with $P_{\rm e}< 0$ and $P_{\rm N}> 0$.\cite{Price71,Hafner87}
\end{namelist}

\section{\label{sec:tensor}Electron and nuclear stress tensors}
In the present section, we derive formulae of stress tensors
by the same arguments that were advanced in the preceding section 
treating the pressure in an electron-nucleus mixture.
In the electron-nucleus mixture where the electrons and nuclei are 
confined in the same volume $V$ (System-I), Nielsen and Martin [NM] \cite{NM85} proved that 
the stress intrinsic to the system is given by a sum of the electron and 
nuclear parts:
\begin{equation}\label{e:Tstress0}
T_{\mu\nu}=T_{\mu\nu}^{\rm ele}+T_{\mu\nu}^{\rm nuc}
\end{equation}
with the definitions of the electron and nuclear stress tensors by
\begin{eqnarray}
T_{\mu\nu}^{\rm ele}&=&-\sum_i\left\langle\frac{\hat p_{i\mu}\hat p_{i\nu}}{m}-\hat{r}_{i\mu}\nabla_{i\nu}\hat{U}\right\rangle\\
T_{\mu\nu}^{\rm nuc}&=&-\!\sum_\ell\left\langle\frac{\hat P_{\ell\mu}\hat P_{\ell\nu}}{M}\!-\!\hat{R}_{\ell\mu}\nabla_{\ell\nu}\hat{U}\right\rangle
=-Nk_{\rm B}T\delta_{\mu\nu}\!-\!\sum_\ell\left\langle\hat{R}_{\ell\mu}{\hat F}_{\ell\nu}\right\rangle\,, \label{e:Tnuc}  
\end{eqnarray}
respectively. In the framework of the DF theory, 
the electron stress tensor is written in the form:
\begin{eqnarray}
T_{\mu\nu}^{\rm ele}
&=&-\sum_k\left\langle f(\epsilon_k)\left(\phi_k\left|\frac{\hat p_{\mu}\hat p_{\nu}}{m}\right|\phi_k\right)\right\rangle_{\rm N}
-\frac12\int \!\!\langle\rho({\bf r})\rho({\bf r}')\rangle_{\rm N}\!\frac{({\bf r}-{\bf r}')_\mu({\bf r}-{\bf r}')_\nu}
{|{\bf r}-{\bf r}'|^3}d{\bf r}d{\bf r}'\nonumber \\
\hspace{-2cm}&&+\int \langle\sigma_{\mu\nu}^{\rm xc}\rangle_{\rm N}d{\bf r}+\sum_\ell \langle{R}_{\ell\mu}F_{\ell\nu}\rangle_{\rm N}\;, 
\label{e:Telc2}
\end{eqnarray} 
where the bracket with N suffix indicates the configurational average of $A(\{{\bf R}_\alpha\})$ defined by
\begin{equation}\label{e:avN}
\langle A\rangle_{\rm N}\equiv\frac1\Xi\int d{\bf R}^N\exp(-\beta {\cal F}_{\rm e}[n;\{{\bf R}_\alpha\}\,]\,) A(\{{\bf R}_\alpha\})
\end{equation}
with
\begin{eqnarray}
\Xi\equiv \int d{\bf R}^N\exp(-\beta {\cal F}_{\rm e}[n;\{{\bf R}_\alpha\}\,]\,)\nonumber\,.
\end{eqnarray}
Here, the exchange-correlation 
stress tensor ${\sigma}^{\rm xc}$ is defined in the DF theory as 
\begin{equation}\label{e:SxcDef}
\nabla\cdot {\sigma}^{\rm xc}\equiv -n({\bf r})\nabla 
{\delta {\cal F}_{\rm xc}\over\delta n({\bf r})}\,,
\end{equation} 
in terms of the exchange-correlation part ${\cal F}_{\rm xc}$ of the free-energy 
for electrons in the fixed nuclei, and $f(\epsilon_k)$ denotes the occupation probability for a state $\epsilon_k$ determined by the Kohn-Sham equation with its wave function $\phi_k$. Also, the electrostatic energy $E_{\rm es}[n]$ of the nucleus-electron mixture is written as
\begin{eqnarray}
\hspace{-1cm}E_{\rm es}[n]&\equiv &\frac{e^2}2\!\!\int {
 n({\bf r})n({\bf r'})\over |{\bf r}-{\bf r'}|}d{\bf r}d{\bf r'}
-\sum_{\ell=1}^N\!\int {Z e^2 n({\bf r}')\over |{\bf r}'-
{\bf R}_{\ell}| } d{\bf 
r}'+\frac12\sum_{\alpha\neq\beta}{Z^2 e^2 
\over |{\bf R}_{\alpha}-{\bf R}_{\beta}|} \label{e:EesA}\\
&\equiv &E_{\rm ee} +E_{\rm en} +E_{\rm nn}
=\frac12\int \frac{\rho({\bf r})\rho({\bf r}')}{|{\bf r}-{\bf r}'|}d{\bf r}d{\bf r}'
=E_{\rm es}[n,\tilde n_{\rm I}]\, \label{e:EesB}
\end{eqnarray}
with $\rho({\bf r})\equiv -en({\bf r})+Ze\tilde n_{\rm I}({\bf r})$ and 
$\tilde n_{\rm I}({\bf r})\equiv \sum_\ell \delta({\bf r}-{\bf R}_{\ell})$. 
In the derivation of (\ref{e:Telc2}), 
we have used the relation:
\begin{eqnarray}\label{e:TEes1}
\hspace{-1cm}\frac12\!\int \!\!\rho({\bf r})\rho({\bf r}')\frac{({\bf r}-{\bf r}')_\mu({\bf r}-{\bf r}')_\nu}
{|{\bf r}-{\bf r}'|^3}d{\bf r}d{\bf r}'&=&-\!\!\int n({\bf r}){r}_\mu\nabla_\nu {\delta E_{\rm es}\over\delta n({\bf r})}d{\bf r}-\!\!\sum_\ell {R}_{\ell\mu}\nabla_{\ell\nu }E_{\rm es}\;. 
\end{eqnarray}
Therefore, by the combined use of (\ref{e:Tnuc}) and (\ref{e:Telc2}), 
the total stress tensor $T_{\mu\nu}$ is represented as
\begin{eqnarray}
T_{\mu\nu}&=&\!-Nk_{\rm B}T\delta_{\mu\nu}-\sum_k\left\langle f(\epsilon_k)\left(\phi_k\left|\frac{\hat p_{\mu}\hat p_{\nu}}{m}\right|\phi_k\right)\right\rangle_{\rm N}\nonumber \\
&&-\frac12\int \!\langle\rho({\bf r})\rho({\bf r}')\rangle_{\rm N}\frac{({\bf r}-{\bf r}')_\mu({\bf r}-{\bf r}')_\nu}
{|{\bf r}-{\bf r}'|^3}d{\bf r}d{\bf r}'+\int \langle\sigma_{\mu\nu}^{\rm xc}\rangle_{\rm N}d{\bf r}\,, 
\label{e:TstressD}
\end{eqnarray}
which is essentially identical with the expression of the stress tensor given 
by NM if the nuclear kinetic term is neglected.

On the other hand, in the system where only the nuclei are confined in the volume $V$ (System II), 
the total stress tensor is given by
\begin{eqnarray}
T_{\mu\nu}&=&T_{\mu\nu}^{\rm nuc}=-Nk_{\rm B}T\delta_{\mu\nu}-\sum_\ell\left\langle{R}_{\ell\mu}F_{\ell\nu}\right\rangle_{\rm N} \label{e:TstressII}  
\end{eqnarray}
which leads to the identical expression to (\ref{e:TstressD}), since the electron stress tensor $T_{\mu\nu}^{\rm ele}$ becomes zero in this system.
On the basis of System-II with the infinite nuclear mass, Ziesche {\it et al.} \cite{ZiescheGN88} and Dal Corso and Resta\cite{Corso94} have proposed to define the stress tensor in terms of the nuclear forces. 
In fact, Dal Corso and Resta 
have used the relation $T_{\mu\nu}^{\rm ele}=0$ in their derivation of 
(\ref{e:TstressD}) from (\ref{e:TstressII}).
In spite of (\ref{e:TstressII}) leading to the same expression to (\ref{e:TstressD}), 
these formulae yield different results with each other for each system, since 
System-I is a homogeneous system while System-II is an inhomogeneous system in 
general.
In a metallic system under high pressure, the electron stress tensor $T_{\mu\nu}^{\rm ele}$ plays 
an important role compared with the nuclear stress tensor $T_{\mu\nu}^{\rm nuc}$.

\section{\label{sec:def}Definitions of electron pressure}
As mentioned in \S\ref{s:vir}, the virial theorem shows that the total pressure 
of an electron-nucleus mixture is given by a sum of the nuclear and electron 
pressures; the concept of the electron pressure is clearly defined  
by using the wall potential ${U}_{\rm w}^{\rm ele}$ 
for the electrons in the form of (\ref{e:Pe2}), which is applicable to this mixture without use of the adiabatic approximation. 
On the other hand, many types of definitions for the electron pressure have been proposed up to date within the {\it adiabatic} approximation.
At the first, the thermodynamic electron pressure is defined 
on the basis of the thermodynamic relation\cite{ChiharaUn} 
\begin{equation} \label{e:ThermPe}
\hspace{-6cm}({\rm I})\qquad\tilde P_{\rm e}\equiv -\left.\frac{\partial {\cal F}_{\rm e}}{\partial V}\right|_{TN_{\rm e}\{{\bf R}_\alpha\}}\,,
\end{equation} 
by introducing the electron free-energy ${\cal F}_{\rm e}[n;\{{\bf R}_\alpha\}]$ 
for the electrons in the electron-nucleus mixture confined in the volume $V$: the nuclei are assumed as classical particles, which produce an external potential for electrons in the system. 
This free energy, ${\cal F}_{\rm e}[n;\{{\bf R}_\alpha\}]$, is defined 
by the electron part of the Hamiltonian for this mixture in the form 
\cite{Chihara}
\begin{equation}\label{e:Felec}
{\cal F}_{\rm e}[n;\{{\bf R}_\alpha\}]\equiv -k_{\rm B}T\ln {\rm Tr}_{\rm e} 
\exp[-\beta(\hat T_{\rm e}+{\hat U})]\,,
\end{equation}
where ${\rm Tr}_{\rm e}$ is the trace operator which is taken over a 
complete set of states of the electrons in the system; 
(\ref{e:Felec}) is the free energy of the electrons 
under the external potential caused by the nuclei 
fixed at $\{{\bf R}_\alpha\}$. 
The DF theory provides an exact expression 
for the free energy of the electrons 
in this system in the form
\begin{equation}
{\cal F}_{\rm e}^{\rm DF}[n;\{{\bf R}_\alpha\}]\equiv 
{\cal F}_0+E_{\rm es}[n,{\tilde n}_{\rm I}]
+{\cal F}_{\rm xc}[n,{\tilde n}_{\rm I}]\,
\end{equation}
with use of ${\cal F}_0$, the free energy of a noninteracting electron gas, 
and the exchange-correlation contribution ${\cal F}_{\rm xc}$ to the free energy.

In contrast to the above macroscopic definitions of the electron pressure, we can define the pressure in a microscopic way.
In this spirit, Bader \cite{Bader80,SrebBader75,BaderAus} introduced the microscopic electron pressure tensor 
$\tilde{\bf\textsf P}_{\rm e}$ by treating 
an electron-gas in the fixed nuclei as an interacting many-body system in the form:
\begin{equation}
\hspace{-8cm}({\rm II})\qquad\nabla\cdot \tilde{\bf\textsf P}_{\rm e}\equiv 
{\bf F}({\bf r})\,,
\end{equation}
where the force density ${\bf F}({\bf r})$ is defined by use of the binary correlation 
function $\Gamma ^{(2)}({\bf r},{\bf r}')$ as
\begin{equation}
\hspace{-2cm}{\bf F}({\bf r})\equiv n({\bf r})\sum_\alpha \nabla \frac{Ze^2}{|{\bf r}-{\bf R}_\alpha |}
-2\int d{\bf r}'\nabla \frac{e^2}{|{\bf r}-{\bf r}' |}\Gamma ^{(2)}({\bf r},{\bf r}')\,.
\end{equation}
In a similar way, More \cite{More79,More85} defined the electron pressure tensor based on the fact that
the electron gas is in hydrostatic equilibrium at each point in the electron-nucleus mixture 
with the pressure variations by electrical forces as is written in the following form:
\begin{equation}\label{e:More}
({\rm III})\qquad\nabla\cdot \tilde{\bf\textsf P}_{\rm e}\equiv -n({\bf r})\nabla \!
\left[-\sum_{\alpha=1}^N\!{Z e^2\over |{\bf r}-
{\bf R}_{\alpha}| }+\int {
e^2 n({\bf r'})\over |{\bf r}-{\bf r'}|}d{\bf r'}\right]
=-n({\bf r})\nabla {\delta E_{\rm es}\over\delta n({\bf r})}\,.
\end{equation}
The same definition was introduced by Bartolotti and Parr \cite{Barto80} from a more general point of view.
On the basis of the DF theory, the microscopic electron pressure tensor 
$\tilde{\bf\textsf P}_{\rm e}$ defined by (\ref{e:More}) is shown to be written 
explicitly by the sum of the electron kinetic tensor 
${\bf\textsf P}_{\rm K}^{\rm DF}$ and the exchange-correlation pressure tensor ${\bf\textsf P}_{\rm xc}^{\rm DF}$:\cite{Chihara01}
\begin{equation}
\tilde{\bf\textsf P}_{\rm e}\equiv {\bf\textsf P}_{\rm K}^{\rm DF}+{\bf\textsf P}_{\rm xc}^{\rm DF}\,,
\end{equation}
where ${\bf\textsf P}_{\rm K}^{\rm DF}$ is defined in terms of the wave functions $\phi_i$ for the Kohn-Sham equation as
\begin{equation}\label{e:Pktns}
{\bf\textsf P}_{\rm K}^{\rm DF}\equiv \sum_i f(\epsilon_i)\frac{-\hbar^2}{4m}\left[
\phi_i\nabla_\mu\nabla_\nu\phi_i^*+\phi_i^*\nabla_\mu\nabla_\nu\phi_i
-\nabla_\mu\phi_i\nabla_\nu\phi_i^*-\nabla_\mu\phi_i^*\nabla_\nu\phi_i \right]\,,
\end{equation}
and the exchange-correlation contribution ${\bf\textsf P}_{\rm xc}^{\rm DF}$ is defined in the DF theory as 
\begin{equation}\label{e:PxcDef}
\nabla\cdot {\bf\textsf P}_{\rm xc}^{\rm DF}\equiv n({\bf r})\nabla 
{\delta {\cal F}_{\rm xc}\over\delta n({\bf r})}\,.
\end{equation} 
In this definition, the thermodynamic pressure is given by the following surface integral over the surface of the system:
\begin{equation}
3\tilde P_{\rm e}V=\oint_{\partial V}{\bf r}\cdot\tilde{\bf\textsf P}_{\rm e}\cdot d{\bf S}\;.
\end{equation}
which can be transformed in the volume-integral form:\cite{Chihara01}
\begin{eqnarray}
\hspace{-1.5cm}3\tilde P_{\rm e}V=2T_s+E_{\rm es}+\int_V {\rm tr} {\bf P}_{\rm xc}^{\rm DF} d{\bf r}
-\sum_\alpha {\bf R}_{\alpha}\!\cdot\!{\bf F}_\alpha
\label{e:DFeos} \,, 
\end{eqnarray} 
in terms of the kinetic-energy functional $T_s$, where the operator ${\rm tr}$ means to take the sum of 
the diagonal terms of tensor.
In the same scheme, another form of microscopic electron pressure tensor was introduced by 
Nielsen, Martin, Ziesche,\cite{NM85,ZiescheGN88} and  Folland \cite{Folland86} with use of the Maxwell pressure tensor ${\bf\textsf P}^{\rm M}$
\begin{equation}\label{e:Pmaxwell}
\hspace{-6.5cm}({\rm IV})\qquad\tilde{\bf\textsf P}^{\rm M}_{\rm e}\equiv {\bf\textsf P}_{\rm K}^{\rm DF}+{\bf\textsf P}_{\rm xc}^{\rm DF}+{\bf\textsf P}^{\rm M}\,.
\end{equation}

On the other hand, the electron pressure is determined  by the formula based on the force theorem:\cite{ChiharaUn,MackAnd75,Skriver84}
\begin{equation}
\hspace{-2.2cm}({\rm V})\qquad 3{\tilde P}_{\rm e}V=-3V\sum_i f(\epsilon_i)\left.\frac{d\epsilon_i}{dV}\right |_{U_{\rm self}}
+\oint_{\partial V}
{\bf r}\cdot{\bf\textsf P}_{\rm xc}^{\rm DF}\cdot d{\bf S}\,,
\end{equation}
where the volume derivative of the Kohn-Sham eigenvalue $\epsilon_i$ 
is performed under the self-consistent potential $U_{\rm self}$ being fixed.
From a different point of view from the above pressure definitions, 
Janak \cite{Janak74} and Ziesche {\it et al},\cite{ZiescheGN88} made a definition of the electron pressure 
with use of the total force on the nuclei in the form
\begin{equation}\label{e:PeNucV}
\hspace{-6.5cm}({\rm VI})\qquad3\tilde{P}_{\rm e}V\equiv \sum_\alpha{\bf R}_\alpha\cdot{\bf F}_\alpha\,. 
\end{equation}
 
In spite of different expressions, it can be shown that all definitions of 
the electron pressure, (I)-(V), except (VI), lead to the same pressure formula 
for the electrons confined in the volume $V$ 
 (see also Appendix~\ref{append1}) as follows:\cite{Chihara01,Bader80,BaderAus}$^,$\cite{ChiharaUn}
\begin{equation}\label{e:PeI}
3\tilde P_{\rm e} V=2 \langle\hat T_{\rm e}\rangle_{\rm e}+\langle\hat{U}\rangle_{\rm e}
-\sum_\alpha\langle\hat{\bf R}_\alpha\cdot\hat{\bf F}_\alpha\rangle_{\rm e}
=\oint_{\partial V}{\bf r}\cdot\tilde{\bf\textsf P}_{\rm e}\cdot d{\bf S}\,.
\end{equation}
It should be noticed that the pressure defined by (\ref{e:Pmaxwell}) with use of 
the Maxwell pressure tensor can play the role of the electron pressure 
only under the condition that the relation, $\oint_{\partial V}
{\bf r}\cdot{\bf\textsf P}^{\rm M}\cdot d{\bf S}=0$, is fulfilled on the surface $\partial V$ of the volume $V$.
In fact, many investigators\cite{Liberman71,Pettifor76,LegPerrot01}$^,$\cite{Skriver84} use a surface-integral expression (\ref{e:PeI}) for calculating the electron pressure in solids.
On the other hand, the definition (VI) leads to the relation \cite{Janak74,ZiescheGN88} 
\begin{equation}\label{e:PevirNuc}
3\tilde P_{\rm e} V
=\sum_\alpha{\bf R}_\alpha\cdot{\bf F}_\alpha
=\sum_\alpha\langle\hat{\bf R}_\alpha\cdot\hat{\bf F}_\alpha\rangle_{\rm e}
=2\langle \hat T_{\rm e}\rangle_{\rm e}+\langle\hat{U}\rangle_{\rm e}\,,
\end{equation}
with use of the adiabatic approximation and the Hellmann-Feynman theorem.
Although the right-hand side of (\ref{e:PevirNuc}) has an appropriate form of the virial theorem consisting of two times of the electron kinetic energy and the potential energy, Eq.~(\ref{e:PevirNuc}) cannot be taken as providing the electron pressure; this equation itself means $P_{\rm e}=0$ because of (\ref{e:Pe2}) in Sec.\ref{s:vir}.
It is rather appropriate to consider that the pressure defined by (\ref{e:PeNucV}) provides the total pressure for another 
inhomogeneous system (System-II), that is, the electron-nucleus mixture 
where only the nuclei are confined in the volume $V$ in the case where the nuclear kinetic pressure can be neglected (see also Appendix \ref{append2}).
In some works\cite{R69,W79}, the electron pressure is defined by the variation  of the electron internal energy per unit cell with respect to the cell volume $\Omega$ 
under the fixed entropy $S$ and total electron number $N_{\rm e}$: 
\begin{equation}\label{e:omegaP}
\hspace{-7cm}({\rm VII})\qquad\tilde P_{\rm e}\equiv -\left.\frac{\partial 
{E}_{\rm e}^{\Omega}}{\partial \Omega}\right|_{SN_{\rm e}}\,.
\end{equation} 
This definition also leads to the relation $3\tilde P_{\rm e}V
=2\langle \hat T_{\rm e}\rangle+\langle\hat{U}\rangle$, 
since the nuclear coordinates are not fixed in the $\Omega$-derivative of this definition (see Appendix~\ref{append1}). Note that the pressure (\ref{e:omegaP}) is not the electron pressure, but provides the total pressure for an electron-nucleus mixture where the nuclear mass is taken to be infinite, as is shown in (\ref{e:Om}).

At this point, it is important to notice that the microscopic pressure tensor $\tilde{\bf\textsf P}^{\rm M}_{\rm e}$ defined by (\ref{e:Pmaxwell}) 
can provide a pressure formula for an arbitrary subspace with a volume $\Omega$ in the system as follows:
\begin{eqnarray}
\hspace{-1.5cm}3\tilde P_{\rm e}\Omega=\oint_{\partial\Omega}{\bf r}\cdot\tilde{\bf\textsf P}^{\rm M}_{\rm e}\cdot d{\bf S} 
&=&\int_\Omega {\rm tr} \tilde{\bf\textsf P}^{\rm M}_{\rm e}d{\bf r}
-\sum_{\alpha\in\Omega} {\bf R}_{\alpha}\!\cdot\!{\bf F}_\alpha
=\oint_{\partial\Omega}{\bf r}\cdot\tilde{\bf\textsf P}_{\rm e}\cdot d{\bf S}\\
&=&2T_s[n]_\Omega+E_{\rm es}(\Omega)+\int_\Omega {\rm tr} {\bf\textsf P}_{\rm xc}d{\bf r}
-\sum_{\alpha\in\Omega} {\bf R}_{\alpha}\!\cdot\!{\bf F}_{\alpha} \label{e:trPxc}\,,
\end{eqnarray} 
if $\tilde{\bf\textsf P}^{\rm M}_{\rm e}$ is constant on the surface $\partial\Omega$ of a volume $\Omega$ and $\oint_{\partial\Omega}
{\bf r}\cdot{\bf\textsf P}^{\rm M}\cdot d{\bf S}=0$.
Here, the electrostatic energy $E_{\rm es}(\Omega)$ contained in the subspace $\Omega$ is defined as
\begin{eqnarray}
E_{\rm es}(\Omega)&\equiv&-\int_\Omega \rho({\bf r}){\bf r}\cdot\nabla \phi({\bf r})d{\bf r}\\
&=&\frac1{8\pi}\int_\Omega |{\bf E}|^2d{\bf r}-\oint_{\partial\Omega}
{\bf r}\cdot{\bf\textsf P}^{\rm M}\cdot d{\bf S} \label{e:EesT2}
\end{eqnarray}
with use of the electric field ${\bf E}=-\nabla \phi$ and the electric potential
\begin{equation}\label{e:potP2}
e\phi ({\bf r})\equiv \int {
e^2 n({\bf r'})\over |{\bf r}-{\bf r'}|}d{\bf r'}
-\sum_{\alpha=1}^N\!{Z e^2\over |{\bf r}-
{\bf R}_{\alpha}| }={\delta E_{\rm es}\over\delta n({\bf r})}\,.
\end{equation}
In the above, it is important to recognize that the electrostatic energy $E_{\rm es}(\Omega)$ 
contained in a subspace $\Omega$ is not always equal to $\int_\Omega |{\bf E}|^2d{\bf r}/8\pi$, 
as was defined in some works.\cite{More79}
Also, the kinetic energy $T_s[n]_\Omega$ involved in the volume $\Omega$ is written from (\ref{e:Pktns}) as \cite{More85,Chihara01} 
\begin{eqnarray}
2T_s[n]_\Omega&\equiv& \int_\Omega{\rm tr} {\bf\textsf P}_{\rm K}^{\rm DF}d{\bf r}\\
&=&\sum_i f(\epsilon_i)\int_\Omega\left[ \phi_i^*\frac{-\hbar^2}{2m}\nabla^2\phi_i +\frac{1}{2m}\left|\frac{\hbar}i \nabla\phi_i\right|^2\right]d{\bf r} \label{e:TsO}\\
&=&2\sum_i f(\epsilon_i)\int_\Omega \phi_i^*\frac{-\hbar^2}{2m}\nabla^2\phi_i 
d{\bf r}+\frac{\hbar^2}{4m}\int_\Omega\nabla^2n({\bf r})d{\bf r}\,. 
\end{eqnarray}

In a similar way with use of the Maxwell pressure tensor, the electron part of 
the stress tensor is given for a subspace $\Omega$ of the system by the formula
\begin{equation}
\tilde{\bf\textsf T}^{\rm ele}\Omega=-\int_\Omega\tilde{\bf\textsf P}^{\rm M}_{\rm e}d{\bf r}
+\sum_{\ell\in\Omega}{\bf R}_{\ell}\circ{\bf F}_{\ell}
=-\oint_{\partial\Omega}{\bf r}\circ\tilde{\bf\textsf P}^{\rm M}_{\rm e}\!\!\cdot d{\bf S}
=-\oint_{\partial\Omega}{\bf r}\circ\tilde{\bf\textsf P}_{\rm e}\!\cdot d{\bf S}\,,
\end{equation}          
with the symbol $\circ$ denoting a dyadic product of two vectors.

\section{\label{sec:eos}Equation of states for liquid metals} 
In \S\ref{s:vir}, we have proven that the pressure of an electron-nucleus mixture 
is expressed by a sum of the electron and nuclear pressures, which are clearly defined 
in the virial forms. 
In the present section, we derive a virial pressure formula for simple liquid metals 
based on this fact, and the zero pressure condition is examined numerically for 
this pressure formula.
\subsection{\label{sec:deriv}Derivation of a virial pressure formula}
The standard virial pressure formula for a simple metal as an electron-ion mixture is written in a complicated 
form to calculate owing to the volume derivative of 
the effective ion-ion interaction in the formula.\cite{Hansen86} In contrast to this standard expression, 
the virial theorem described in \S\ref{s:vir} provides the equation of states (EOS) in a simple form:

\begin{eqnarray}
\hspace{-1cm} 3PV&=&3(P_{\rm e}+P_{\rm N})V\equiv 2\langle \hat T_{\rm e}\rangle-\sum_i\langle\hat{\bf r}_i\cdot\nabla_i\hat{U}\rangle
+2\langle \hat T_{\rm N}\rangle-\sum_\alpha\langle\hat{\bf R}_\alpha\cdot\nabla_\alpha\hat{U}\rangle \label{e:Ptherm2} \\
\hspace{-1cm}&= &3P_{\rm e}V+3Nk_{\rm B}T+\sum_\alpha \left<{\bf R}_{\alpha}\!\cdot\!{\bf F}_\alpha\right>_{\rm N}\,,\label{e:Ptherm3}
\end{eqnarray}
by regarding liquid metals as a mixture of electrons and nuclei.
In a simple liquid metal, the interaction among ions  with the Coulomb potential $v_{\rm II}^c(Q)$ can be approximated 
as a sum of binary interaction $v_{\rm II}^{\rm eff}(r)$, which is defined 
in the quantum hypernetted chain (QHNC) theory \cite{Chihara89} by 
\begin{equation}\label{eq:veff}
\beta v_{\rm II}^{\rm eff}(Q)\equiv\beta v_{\rm II}^c(Q)
 -{|C_{\rm eI}(Q)|^2 n_0^{\rm e} \chi_Q^0\over
 1+\beta n_0^{\rm e} v_{ee}^c(Q)[1-G(Q)] \chi_Q^0 }\;.
\end{equation}
Here, $C_{\rm eI}(Q)$ denotes the electron-ion direct correlation 
function, which plays the role of a pseudopotential $w(Q)=-C_{\rm eI}(Q)/\beta$, 
and $\chi_Q^0$, the density-density response function of a noninteracting electron gas, 
with $G(Q)$ being the local-field correction.
Therefore, with use of this two-body interaction, the EOS is written as follows:
\begin{eqnarray}\label{e:Leos1}
3PV&=&3P_{e}V+3Nk_{\rm B}T
-N\frac12 n_0^{\rm I}
\int g_{\rm II}(r){\bf r}\nabla v_{\rm II}^{\rm eff}(r)d{\bf r}\label{e:fleos1}\,.
\end{eqnarray}

At the next step, we proceed to derive an approximate representation of 
the electron pressure $P_{\rm e}$ in (\ref{e:Leos1}) on the basis of the spherical cell model. 
By taking a Wigner-Seitz sphere with a radius $R$, the electron pressure 
can be represented by the surface integral over the surface $S$ in the form:\cite{Chihara01} 
\begin{equation}\label{e:Pjell}
P_{\rm e}=\frac1{3\Omega}\oint_S{\bf r}\cdot\left({\bf P}_{\rm K}^{\rm DF}+{\bf P}_{\rm xc}^{\rm DF}\right)\cdot d{\bf S}=P_{\rm k}(R)+P_{\rm xc}(R)\;.
\end{equation}
In the above, the kinetic pressure $P_{\rm k}(R)$ is given in units of hartree in the form:\cite{LegPerrot01}
\begin{equation}\label{e:Psurf}
 P_{\rm k}(R)=\frac1{8\pi}\sum_{k\ell}f(\epsilon_{k\ell}\!-\!\bar{\mu})
|\phi_{k\ell}(R)|^2\left[\frac{(D_{k\ell}\!-\!\ell)(D_{k\ell}\!+\!\ell+1)}{R^2}
+2\{\epsilon_{k\ell}-V_{\rm eff}(R)\}\right]\,,
\end{equation}
with
\begin{equation}
D_{k\ell}\equiv R\frac1{\phi_{k\ell}}\left.\frac{d\phi_{k\ell}}{dr}\right|_R\,,
\end{equation}
and $\bar{\mu}$ being the chemical potential.
In the above, $\phi_{k\ell}(r)$ denotes the radial part of the wave function 
for a state $\epsilon_{k\ell}$ of angular momentum $\ell$ satisfying 
\begin{equation}
\frac{d^2(r\phi_{k\ell})}{dr^2}+\left[2(\epsilon_{k\ell}-V_{\rm eff})
-\frac{\ell(\ell+1)}{r^2} \right](r\phi_{k\ell})=0\,.
\end{equation}

In the SJ model,\cite{Shore91} the whole crystal with dipole layer $D$ at its surface 
is considered as an assembly of Wigner-Seitz cells with dipole-layer surface.
As a result, by taking the zero of a potential to be at the vacuum level, 
the effective potential $V_{\rm eff}$ in the Wigner-Seitz sphere is written as
\begin{equation}
V_{\rm eff}({\bf r})=v_{\rm eN}+\int v_{\rm ee}({\bf r}\!-\!{\bf r}')n({\bf r}')d{\bf r}'
+\mu_{\rm xc}({\bf r})-D\theta(R-r)
\end{equation}
with $\theta(x)$ being the step function and $\mu_{\rm xc}({\bf r})\equiv {\delta {\cal F}_{\rm xc}/\delta n({\bf r})}$.
With a choice of the zero of the potential to be $V_{\rm eff}(R^-)$, 
the electron pressure at the surface $S$ takes a form:
\begin{equation}\label{e:Psurf2}
P_{\rm k}(R)=\!-\!Dn(R)+\frac{1}{24\Omega\pi}\sum_{k\ell}f(\epsilon_{k\ell}^0\!-\epsilon_{\rm F})
|\phi_{k\ell}(R)|^2\left[\frac{(D_{k\ell}\!-\!\ell)(D_{k\ell}\!+\!\ell+1)}{R^2}
+2\epsilon_{k\ell}^0\right]\,.
\end{equation}
In the above, $\epsilon_{k\ell}^0$ and $\epsilon_{\rm F}$ indicate an energy level 
and Fermi energy with respect to this new origin, respectively, and the pressure at the surface is defined as $P_{\rm k}(R)\equiv P_{\rm k}(R^+)$.

When the wave functions of electrons become plane waves and produce a constant density $n$ near the boundary of the Wigner-Seitz sphere, 
the above equation leads to an approximate expression for the electron pressure as follows:
\begin{equation}\label{e:PDn}
P_{\rm e}(n)=-\!Dn+n^2\frac{d\epsilon_{\rm J}}{dn}\,,
\end{equation}
in terms of the energy $\epsilon_{\rm J}$ per one electron in the jellium model.
\begin{table}[b] 
\caption{The nuclear pressures (GPa) for alkaline liquid metals in the vacuum 
calculated by the QHNC method and the Ashcroft model potential (Rb). The nuclear kinetic and virial pressures are 
also shown together with the electron pressure given by 
$P_{\rm e}^{\rm jell}=\left.n_0^2{d\epsilon_{\rm J}}/{dn}\right|_{n_0}$ (the 
jellium model). 
The nuclear pressures calculated by Kumaravadivel \cite{Kuma83} deviate 
significantly from the zero pressure condition compared with the QHNC results.
}
\vspace{0.2cm} 
\begin{tabular}{ll|cc|cc||cc|cc}\hline
      & $r_s^0$ & $n_{\rm I}k_{\rm B}T$ & nuclear virial &   $P_{\rm N}$ & 
$P_{\rm N}$\cite{Kuma83} &$\left.n_0^2{d\epsilon_{\rm J}}/{dn}\right|_{n_0}$ \\ \hline   
Li    &	 3.308  &  0.2888 &-0.2871   & 0.0017 &-1.7  &  2.70 \\
Na    &	 4.046  &  0.1253  & 0.0362    & 0.1615   &-0.24  & 0.024 \\
Rb    &	 5.388  &  0.0445  & 0.0734    & 0.1179   & -0.89& -0.43 \\
      &5.388 &  0.0445  & -0.0458   & -0.0013 &      & -0.43 \\
\end{tabular}
\label{tbl:2}
\end{table} 

In the jellium model, the electron pressure does not 
satisfy the condition  $P_{\rm e}=0$ at equilibrium density $n_0$ in the vacuum. 
In contrast to this model the condition  $P_{\rm e}=0$ at equilibrium density $n_0$ is fulfilled 
in the stabilized jellium (SJ) model \cite{Perdew90,Shore99} 
(ideal metal model \cite{Shore91,Shore99}), where the finite jellium has 
an infinitesimally thin dipole layer on its surface.
The relation (\ref{e:PDn}) provides the strength of the dipole layer 
at equilibrium [$P_{\rm e}(n_0)=0$] of a metal in the vacuum: 
$D(n_0)=[nd\epsilon_{\rm J}/dn]_{n_0}$, which is identical with the 
result determined from the SJ model. In the SJ model the strength of 
the dipole layer is assumed to be given by the relation:
\begin{equation}
D(n)=\frac{n}{n_0}D(n_0)\,.
\end{equation}
With use of this approximation the electron pressure in a liquid metal can 
be determined finally by
\begin{equation}\label{e:Psjell}
P_{\rm e}(n)=n^2\left[\frac{d\epsilon_{\rm J}}{dn}-\left.\frac{d\epsilon_{\rm J}}{dn}\right|_{n_0} \right]\,.
\end{equation}

On the other hand, Nieminen and Hodge \cite{NiemHodg76} showed that application of 
the relation (\ref{e:Psurf}) to a free electron of plane waves leads to the 
expression for the pressure in terms of the internal chemical potential $\bar{\mu}$:
\begin{equation}\label{e:Pmu}
P_{\rm e}(n)=n(\bar{\mu}-\epsilon_{\rm J})\,,
\end{equation}
which can be also derived by the TF theory with the Weizsacher correction.\cite{Meyer71,P79,More79,Barto80}
At this point, Eq.~(\ref{e:Pmu}) can be rewritten in the form:
\begin{equation}
P_{\rm e}=n(\bar{\mu}-\mu_{\rm J})+ n^2\frac{d\epsilon_{\rm J}}{dn}
\end{equation}
with $\mu_{\rm J}=d(n\epsilon_{\rm J})/dn$. By comparing this equation with 
(\ref{e:PDn}) we obtain the following relation:
\begin{equation}
D=\mu_{\rm J}-\bar{\mu}\,,
\end{equation}
which leads to the formula for work function $w$:\cite{Heine72} 
$w=D-\mu_{\rm J}=-\bar{\mu}$ with $\mu_{\rm J}$ providing the
internal chemical potential. 
From this equation we see that 
the relation $-D/n=(\bar{\mu}-\mu_{\rm J})/n=-d\epsilon_{\rm J}/dn|_{n_0}$ 
is assumed in the SJ model.

It has been already discussed about the problem 
that the electron pressure for a plasma does not become zero at equilibrium 
in the vacuum: in fact, in constructing the EOS for a plasma by 
the use of the ion-sphere model based on the TF method, a similar correction to (\ref{e:Psjell}) is
introduced as a chemical bonding correction (Barnes correction) to make the electron pressure 
zero at equilibrium in the vacuum by More {\it et al}.\cite{MWYZ88}
In the same spirit, we adopt (\ref{e:Psjell}) as the electron pressure $P_{\rm e}$ in (\ref{e:fleos1}): then, the total pressure for liquid metals is described in the final form: 
\begin{equation}\label{e:fleos2}
P=n^2\left[\frac{d\epsilon_{\rm J}}{dn}-\left.\frac{d\epsilon_{\rm J}}{dn}\right|_{n_0} \right]+n_0^{\rm I}k_{\rm B}T-\frac16 (n_0^{\rm I})^2
\int g_{\rm II}(r){\bf r}\nabla v_{\rm II}^{\rm eff}(r)d{\bf r} \,.
\end{equation}
At this point, it should be noticed that the following relation
\begin{equation}\label{e:Pn0}
n_0^{\rm I}k_{\rm B}T=\frac16 (n_0^{\rm I})^2
\int g_{\rm II}(r){\bf r}\nabla v_{\rm II}^{\rm eff}(r)d{\bf r}
\end{equation}
must be satisfied at equilibrium in the vacuum, where $P_{\rm N}=0$. 

\begin{figure}[b]
\leavevmode
\epsfxsize=9cm
\epsffile{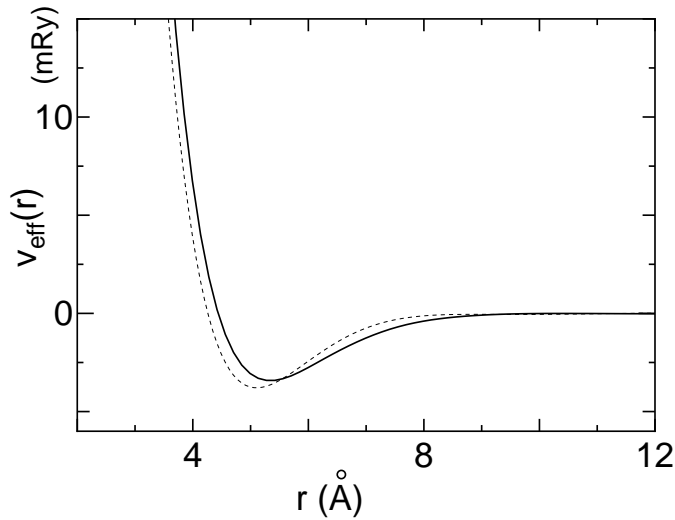}
\caption{The effective ion-ion interactions of Rb (313K) determined by the QHNC equation 
and by the use of Ashcroft model potential ($r_c=1.275$), 
which are shown by the full and dotted curves, respectively. 
}
\label{fig-1}
\end{figure}

\subsection{\label{sec:qhnc}Numerical examination of the QHNC equation by the zero-pressure condition}
In this subsection, we exhibit how the zero-pressure condition, (\ref{e:Pn0}),
 can be used as a check of numerical consistency in the calculation 
by applying it to some alkaline liquid metals at the normal pressure on the 
basis of the QHNC equation\cite{Kamb96} as an example.
Since we can regard liquid metals at the normal pressure as those in the vacuum, 
the pressures determined from (\ref{e:fleos2}) must be zero in this circumstance:
that is, $P_{\rm N}=0$ while $P_{\rm e}=0$ by definition.
In order to examine this relation we calculated the nuclear virial term and nuclear kinetic pressures 
from (\ref{e:fleos2}), which are shown in Table~\ref{tbl:2} for Li, Na and Rb.

In the case of Li, the nuclear virial term becomes -0.2871 GPa, 
which provides almost zero total pressure 0.0017 GPa by canceling out the
nuclear kinetic pressure 0.2888 GPa. Thus, the QHNC equation can be considered to give a reasonable result for Li to fulfill (\ref{e:Pn0}) in comparison with 
-1.7 GPa of Kumaravadivel.\cite{Kuma83}
However, the QHNC equation does not provide 
even negative values of the nuclear virial term for the Na and Rb cases; 
as a result, the nuclear pressure remains to be a large positive value for each case.

\begin{figure}[b]
\leavevmode
\epsfxsize=9cm
\epsffile{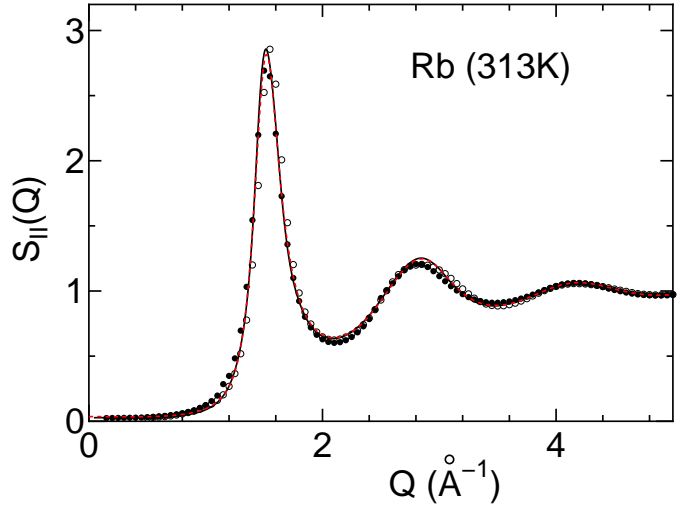}
\caption{The ion-ion structure factors $S_{\rm II}({\bf Q})$ of Rb obtained by use of the effective 
ion-ion interactions based on the QHNC and Ashcroft models, 
which are displayed by the full and dotted curves, respectively.
The full and white circles indicate the experimental results by 
the neutron \cite{Copley77} and X-ray \cite{Waseda80} diffraction methods, respectively.  The Ashcroft model 
with $r_c=1.275$\AA\ determined by the condition $P_{\rm N}=0$ provides an 
indistinguishable structure factor from the QHNC result in a good agreement 
with the experiments. 
}
\label{fig-2}
\end{figure}

To see where this discrepancy comes from, we calculated the pressure of Rb 
from an effective ion-ion interaction determined by the Ashcroft model 
potential with the core parameter 
$r_{\rm c}=1.27$\AA\ \cite{Hansen86} on the basis of (\ref{e:fleos2}): 
the nuclear pressure becomes -0.0013 GPa, which satisfies the condition (\ref{e:Pn0}) 
 to a considerable extent compared to other calculation \cite{Kuma83} shown in Table \ref{tbl:2}.
Inversely, the condition (\ref{e:Pn0}) can be used to determine the Ashcroft 
core parameter $r_{\rm c}$; this condition generates $r_{\rm c}=1.275$\AA\ for Rb case.
In Fig. \ref{fig-1}, the effective ion-ion interactions for a Rb liquid metal obtained from the Ashcroft model with $r_{\rm c}=1.275$\AA\ are plotted 
together with that from the QHNC method.
Although we find a visible difference between these two effective potentials  in this scale, 
these potentials yield almost the same structure factors showing a good 
agreement with experiments as is displayed in Fig. \ref{fig-2}.
In contrast to the Ashcroft model potential, the pseudopotential $-\!C_{\rm eI}(r)/\beta$ 
determined from the QHNC equation has a complicate inner structure in the core region. 
As a consequence, the term $\int C_{\rm eI}^{\rm R}(r)d{\bf r}$ [$C_{\rm eI}^{\rm R}(r)$: the non-Coulomb part] which appears in 
(\ref{e:Pn0}) is difficult to calculate with a sufficient accuracy; 
the discrepancy of the QHNC result for the zero-pressure condition is attributed to this problem in the Rb and Na 
cases in contrast to the case of a Li metal which has a simpler inner core structure in the pseudopotential $C_{\rm eI}$. It is well known that the inner-core structure of a pseudopotential is not significant to determine the effective ion-ion interaction, but is important to calculate the thermodynamical quantities, as the QHNC results indicate.
It is worth noting that the condition (\ref{e:Pn0}) provides a check of numerical consistency in the calculation for determining the effective ion-ion interaction, the structure factor and the thermodynamical quantities for a liquid metal in the vacuum.

\section{\label{sec:dis}Summary and discussion} 

In general, the virial theorem (\ref{e:VirTh}) teaches us that 
the partial pressure associated with each component in a mixture can be uniquely defined by using the wall potential for each component, and the total pressure is represented as a sum of them: this definition is assured by the virial theorem for single particle, (\ref{e:vE}) and (\ref{e:vN}). Also, the partial pressure can be actually measured as an osmotic pressure, for example. Therefore, 
in an electron-nucleus mixture, the electron and nuclear pressures 
are clearly and simply defined as the partial pressures owing to the virial theorem  (\ref{e:VirTh}) as is shown in \S\ref{s:vir}, and 
the thermodynamic pressure of this system is given by a sum of the electron and 
nuclear pressures. 
To be consistent with this fact, the electron pressure in this mixture 
is defined by
\begin{equation}\label{e:PeVir21}
3P_{\rm e}V\equiv 2\langle \hat T_{\rm e}\rangle-\sum_i\langle\hat{\bf r}_i\cdot\nabla_i\hat{U}\rangle
=2\langle T_{\rm e}\rangle_{\rm N}+\langle U_{\rm e}\rangle_{\rm N}-\sum_\alpha\langle{\bf R}_\alpha\cdot{\bf F}_\alpha\rangle_{\rm N} \;,
\end{equation}
which makes a contrast to another definition \cite{Janak74,ZiescheGN88} 
in terms of the forces on  the nuclei
\begin{equation}\label{e:PbF}
3P_{\rm e}V\equiv
\sum_\alpha\langle{\bf R}_\alpha\cdot{\bf F}_\alpha\rangle_{\rm N}
=2\langle T_{\rm e}\rangle_{\rm N}+\langle{U_{\rm e}}\rangle_{\rm N}\;.
\end{equation}
It should be pointed out that the pressure represented by (\ref{e:PbF}) 
is neither the electron pressure nor the total pressure for this system, 
and that this pressure is associated with the total pressure of another inhomogeneous system (System-II as is named in \S\ref{s:vir}), where only the nuclei 
are confined in the volume $V$ and the electron can move over all space under 
the external potential caused by the confined nuclei.
Similar remarks are made about the definition of the stress tensor for 
this mixture as is discussed in \S\ref{sec:tensor}

In \S\ref{sec:def}, we have enumerated several kinds of definitions for the electron pressure, 
which lead to the electron-pressure expression (\ref{e:PeVir21}), where the nuclear virial term is subtracted.
It is important for the calculation of the total pressure by adding 
the nuclear pressure that we should recognize which definition is adopted 
to determine the electron pressure. For example \cite{LegPerrot01} 
as was shown by (\ref{e:PeI}), the electron pressure calculated 
by use of the surface integral expression, 
$3\tilde P_{\rm e}V=\oint_{\partial V}{\bf r}\cdot\tilde{\bf\textsf P}_{\rm e}\cdot d{\bf S}$,
is different from the pressure determined on the basis of the relation, 
$3\tilde P_{\rm e}V=2T_s[n]\!+\!E_{\rm es}\!+\!\int{\rm tr}
{\bf\textsf P}_{\rm xc}^{\rm DF}d{\bf r}$, in that the latter contains 
a part of the nuclear pressure 
$\sum_\alpha{\bf R}_\alpha\cdot{\bf F}_\alpha$, 
which is approximated by the Madelung energy term: $NCZ^2_{\rm out}/4a$ as is given by Janak.\cite{Janak74}

When the nuclei behave as {\it classical} particles, 
the partition function of an electron-nucleus mixture 
with the Hamiltonian (\ref{e:Hen}) is written exactly in the form \cite{Chihara}
\begin{eqnarray}\label{e:ZnP} 
Z_N= \frac{1}{N!h^{3N}}\int\!\! d{\bf R}^N\!\!\int 
d{\bf P}^N\exp\left[-\beta \left(\sum_\alpha \frac{{\bf P}_\alpha^2}{2M}+
{\cal F}_{\rm e}[n;\{{\bf R}_\alpha\}]\right)\right]\,,
\end{eqnarray}
where the electron free energy ${\cal F}_{\rm e}[n;\{{\bf R}_\alpha\}]
$ defined by (\ref{e:Felec}) plays a role of a many-body interaction 
among the nuclei:\cite{ChiharaUn}
\begin{eqnarray}\label{e:forceDE}
{\bf F}_\alpha
&=&-\nabla_{\alpha}\left.{\cal F}_{\rm e}\right|_{TVN_{\rm e}}
=-\nabla_{\alpha}\left.E_{\rm es}\right|_{n({\bf r})}\\
&=&Ze^2\int n({\bf r})\frac{{\bf r}-{\bf R}_\alpha}{|{\bf r}-{\bf R}_\alpha|^3}d{\bf r}
+(Ze)^2\sum_{\alpha\ne \gamma}\frac{{\bf R}_\alpha-{\bf R}_\gamma}{|{\bf R}_\alpha-{\bf R}_\gamma|^3}\label{e:fHF}\,.
\end{eqnarray}
Thus, the pressure of this system is obtained by performing the volume derivative of the free energy $F=-k_{\rm B}T\ln Z_N$ without use of any approximation in the form \cite{Chihara01}
\begin{eqnarray}
3PV=-\left.3V\frac{\partial F}{\partial V}\right|_{TN}
=\left< 
-3V\frac{\partial {\cal F}_{\rm e}}{\partial V}\right>_{\rm N}
+3Nk_{\rm B}T+\left< \sum_\alpha {\bf R}_{\alpha}\!\cdot\!{\bf F}_\alpha\right>_{\rm N}\,,\label{e:DFvir}
\end{eqnarray} 
which is identical with (\ref{e:Ptherm3}); this relates the electron pressure determined in \S\ref{sec:def} under the adiabatic approximation to the total pressure and to that by the virial theorem (\ref{e:Ptherm2}) owing to (\ref{e:ThermPe}). That is, the relation, $P=P_{\rm e}+P_{\rm N}$, is proven {\it without use of the wall potentials} by (\ref{e:DFvir}) in the adiabatic approximation.\cite{ChiharaUn} 
When we approximate ${\cal F}_{\rm e}$ involved in (\ref{e:ZnP}) as
\begin{eqnarray}\label{e:Bapp}
{\cal F}_{\rm e}[n;\{{\bf R}_\alpha\}]\approx U_0(V)
+\frac12\sum_{\alpha\neq \beta}v_{\rm II}^{\rm eff}(|{\bf R}_\alpha-{\bf R}_\beta|)\,,
\end{eqnarray}
and perform the volume derivative of $F$, we obtain the standard virial EOS \cite{Hansen86} for simple liquid metals. 
On the other hand, the virial EOS (\ref{e:fleos2}) is derived by applying the approximation (\ref{e:Bapp}) only to the nuclear virial term of (\ref{e:DFvir}); the first term of (\ref{e:DFvir}) is treated appropriately as the electron pressure in contrast with the standard virial formula,\cite{Hansen86} in which derivation the approximation (\ref{e:Bapp}) is applied to both terms involving ${\cal F}_{\rm e}$ in (\ref{e:DFvir}). As a result, the electron pressure in the exact expression (\ref{e:DFvir}) becomes to be represented by
\begin{equation}
P_{\rm e}\approx (n_0^{\rm I})^2\frac{d U_0(n_0^{\rm I})}{d n_0^{\rm I}} +\frac12 (n_0^{\rm I})^3
\int g_{\rm II}(r)\frac{\partial v_{\rm II}^{\rm eff}(r)}{\partial n_0^{\rm I}}d{\bf r}\,,
\end{equation}
in the standard virial formula for a liquid metal.
Therefore, the EOS (\ref{e:fleos2}) is more accurate than the standard virial formula in spite of its simpler structure. 

A metal in the vacuum is defined as an electron-nucleus mixture without the wall potentials, which confine the electrons ($U_{\rm w}^{\rm ele}$) and the nuclei ($U_{\rm w}^{\rm nuc}$) in a finite volume. From this definition, we can give another important remark that the zero pressure of an electron-nucleus 
mixture in the vacuum is realized by both electron and nuclear pressures 
being zero at the same time: 
this can also be verified directly from thermodynamics as described in Appendix~\ref{append1}. 
In the adiabatic approximation, the electron pressure is given by thermodynamics in the form 
\begin{eqnarray} 
3P_{\rm e}V\equiv-\left\langle 3V\frac{\partial {\cal F}_{\rm e}[n;\{{\bf R}_\alpha\}]}{\partial V}\right\rangle_{\rm N}&=&\langle\{\,2T_{\rm e}+U_{\rm e}
-\sum_\alpha{\bf R}_\alpha\cdot{\bf F}_\alpha \,\}\rangle_{\rm N} \nonumber\\
&=&2\langle \hat T_{\rm e}\rangle-\sum_i\langle\hat{\bf r}_i
\cdot\nabla_i\hat{U}\rangle \,,
\end{eqnarray}
which is zero in the vacuum due to (\ref{e:vE}), since a metal in the vacuum is 
not confined by the wall. Similarly, $P_{\rm N}=0$ due to (\ref{e:vN}).
Therefore, in the first-principles molecular dynamics (MD) performed on liquid metals in the vacuum, for example, the nuclear pressure 
should become zero ($P_{\rm N}=0$); this relation also can be used to check numerical consistency in the MD calculation. 

\appendix
\section{Electron pressure determined by the DF theory}\label{append1} 
The electron pressure can be expressed in terms of the internal energy
of the interacting electrons in the presence of the external potential caused by the fixed nuclei at $\{{\bf R}_\alpha\}$ as follows:\cite{ChiharaUn}
\begin{equation}
\tilde P_{\rm e}\equiv -\left.\frac{\partial {\cal F}_{\rm e}^{\rm DF}}{\partial V}\right|_{TN_{\rm e}\{{\bf R}_\alpha\}}
=-\left.\frac{\partial E^V}{\partial V}\right|_{SN_{\rm e}\{{\bf R}_\alpha\}}\,.\end{equation}
In the DF theory, this electron internal energy is written as
\begin{equation}\label{e:efree1}
E^V[n;\{{\bf R}_\alpha\}]\equiv T_s^V[n]+E_{\rm es}^V[n;\{{\bf R}_\alpha\}]
+{E}_{\rm xc}^V[n]\,.
\end{equation}
Here, the exchange-correlation part of the internal energy is defined by
\begin{equation}
{E}_{\rm xc}^V[n]\equiv {\cal F}_{\rm xc}[n]-T\left.\frac{\partial {\cal F}_{\rm xc}[n]}{\partial T}\right|_{Vn({\bf r})}\,,
\end{equation}
and the kinetic-energy functional $T_s^V[n]$ and the electrostatic energy 
$E_{\rm es}^V[n;\{{\bf R}_\alpha\}]$ of this system are given by (\ref{e:TsO}) and (\ref{e:EesA}), respectively.
With use of the scaled electron density distribution $n_\lambda(r)\equiv 
\lambda^3n(\lambda{\bf r})$, the electron pressure $\tilde P_{\rm e}$ for this system 
can be determined by $\lambda$ derivative in the form:
\begin{equation}
3\tilde P_{\rm e}V=-3V\left.\frac{\partial E^V[n;\{{\bf R}_\alpha\}]}{\partial V}\right|_{N_{\rm e}}
=\left.\lambda \frac{\partial E_\lambda}{\partial \lambda}\right|_{\lambda=1}\,, 
\end{equation}
with
\begin{equation}
E_\lambda\equiv E^{V/\lambda^3}[n_\lambda;\{{\bf R}_\alpha\}]
=T_s^{V/\lambda^3}[n_\lambda]+E_{\rm es}^{V/\lambda^3}[n_\lambda;\{{\bf R}_\alpha\}]+E_{\rm xc}^{V/\lambda^3}[n_\lambda]\,.
\end{equation}
In the above, the scaled density distribution $n_\lambda(r)$ keeps the total electron number as a constant in the
$\lambda$-variation: $N_{\rm e}=\int_Vn({\bf r})d{\bf r}=\int_{V/\lambda^3}n_\lambda({\bf r})d{\bf r}$.
With use of the following relations
\begin{eqnarray}
T_s^{V/\lambda^3}[n_\lambda]&=&\lambda^2 T_s^V[n]\,, \\
E_{\rm es}^{V/\lambda^3}[n_\lambda;\{{\bf R}_\alpha\}]
&=&\lambda E_{\rm es}^V[n;\{\lambda{\bf R}_\alpha\}]\,,
\end{eqnarray}
and the exchange-correlation contribution \cite{Chihara01}
\begin{equation}
 \left.\lambda \frac{\partial E_{\rm xc}^{V/\lambda^3}[n_\lambda]}{\partial
 \lambda}\right|_{\lambda=1}\!\!=-\int_V n({\bf r}){\bf r}\cdot\nabla\left.{\delta {\cal F}_{\rm xc}\over\delta n({\bf r})}\right|_{TV} d{\bf r}+\oint_{\partial V}{\bf r}\cdot{\bf\textsf P}_{\rm xc}^{\rm DF}\cdot d{\bf S}
=\int_V {\rm tr}{\bf\textsf P}_{\rm xc}^{\rm DF}d{\bf r}\,,
\end{equation}
we get finally the expression for the electron pressure in the form:
\begin{eqnarray}
3\tilde P_{\rm e}V&=&\left.\lambda \frac{\partial E_\lambda}{\partial \lambda}\right|_{\lambda=1}
=2T_s^V[n]+E_{\rm es}^V+\int_V {\rm tr} {\bf\textsf P}_{\rm xc}^{\rm DF}d{\bf r}
-\sum {\bf R}_{\alpha}\!\cdot\!{\bf F}_\alpha\\
&=&2T_{\rm e}+U_{\rm e}
-\sum_\alpha{\bf R}_\alpha\cdot{\bf F}_\alpha \label{e:PeTrue}\,,
\end{eqnarray}
in terms of the true kinetic and potential energies 
given by $T_{\rm e}=T_s^V[n]+T_{\rm xc}$ and $U_{\rm e}=E_{\rm es}^V+U_{\rm xc}$, 
respectively. Here, the exchange-correlation contributions to the kinetic and potential energies, $T_{\rm xc}$ and $U_{\rm xc}$, are defined by the relations
\begin{eqnarray}
2T_{\rm xc}+U_{\rm xc}&=&\int_V {\rm tr} {\bf\textsf P}_{\rm xc}^{\rm DF}d{\bf r}\,, \\
T_{\rm xc}+U_{\rm xc}&=&E_{\rm xc}^V\,.
\end{eqnarray}
On the other hand, in the adiabatic approximation, we have an average relation, 
$\langle\hat A\rangle=\langle\langle\hat A\rangle_{\rm e}\rangle_{\rm N}$,  
where the bracket with a suffix e denotes the average over the electron states 
at a fixed nuclear configuration. 
Therefore, with use of this approximation and (\ref{e:PeTrue}), 
the electron pressure (\ref{e:Pe2}) defined by the virial theorem can be 
written as
\begin{eqnarray} 
2\langle \hat T_{\rm e}\rangle-\sum_i\langle\hat{\bf r}_i
\cdot\nabla_i\hat{U}\rangle&=&\langle\{\,2\langle\hat T_{\rm e}\rangle_{\rm e}+\langle\hat U\rangle_{\rm e}
-\sum_\alpha{\bf R}_\alpha\cdot{\bf F}_\alpha\,\}\rangle_{\rm N}\\ 
&=&\langle\{\,2T_{\rm e}+U_{\rm e}
-\sum_\alpha{\bf R}_\alpha\cdot{\bf F}_\alpha \,\}\rangle_{\rm N}
=3\langle \tilde P_{\rm e}\rangle_{\rm N}V\,,
\end{eqnarray}
which is identical with (\ref{e:PeI}). 
It is important to note that the nuclei are fixed at the positions $\{{\bf R}_\alpha\}$ when the volume derivative is performed to determine the electron pressure.

\section{Criticism on the Janak derivation of the pressure formula}\label{append2}
In the DF theory for the electron gas confined in the volume $V$ [System-I], the kinetic 
energy $T_s[n]$ is written as
\begin{eqnarray}
2T_s[n]&=&-\int_V{\bf r}\cdot\nabla\cdot{\bf\textsf P}_{\rm K}^{\rm DF}d{\bf r}+\oint_{\partial V}{\bf r}\cdot{\bf\textsf P}_{\rm K}^{\rm DF}\cdot d{\bf S}
=\int_V {\rm tr} {\bf\textsf P}_{\rm K}^{\rm DF} d{\bf r}  \label{e:TsTr1}\\
&=&\int_V n({\bf r}){\bf r}\cdot\nabla U_{\rm eff}({\bf r})d{\bf r}+\oint_{\partial V}{\bf r}\cdot{\bf\textsf P}_{\rm K}^{\rm DF}\cdot d{\bf S} \label{e:TsTr2} \,,
\end{eqnarray} 
in terms of the kinetic pressure tensor ${\bf\textsf P}_{\rm K}^{\rm DF}$ defined 
by
$\nabla\cdot{\bf\textsf P}_{\rm K}^{\rm DF}
=-n({\bf r})\nabla U_{\rm eff}({\bf r})$.
Also,  
the exchange-correlation pressure tensor ${\bf\textsf P}_{\rm xc}$, which is 
defined by the relation: 
$\nabla\cdot {\bf\textsf P}_{\rm xc}\equiv n({\bf r})\nabla{{\delta E}_{\rm xc}/\delta n({\bf r})}$,\enspace satisfies 
\begin{eqnarray}
\int_V {\rm tr}{\bf\textsf P}_{\rm xc}d{\bf r}&=&-\int_V{\bf r}\cdot\nabla\cdot {\bf\textsf P}_{\rm xc} d{\bf r}+\oint_{\partial V}{\bf r}\cdot{\bf\textsf P}_{\rm xc}\cdot d{\bf S} \label{e:dExcdlam1}\\
&=&-\int_V n({\bf r}){\bf r}\cdot\nabla{\delta {E}_{\rm xc}\over\delta n({\bf r})} d{\bf r}+\oint_{\partial V}{\bf r}\cdot{\bf\textsf P}_{\rm xc}\cdot d{\bf S} \label{e:dExcdlam2} \,.
\end{eqnarray}
Note that Eqs. (\ref{e:TsTr1}) and (\ref{e:dExcdlam1}) are a general relation which must be satisfied for any tensor in the electron gas confined in the volume $V$. Nevertheless, the surface integrals in (\ref{e:TsTr2}) and (\ref{e:dExcdlam2}) are {\it neglected} in Janak's equations,\cite{Janak74} (7) and (10) in the local-density approximation, respectively.

Since the electrostatic energy $E_{\rm es}$, (\ref{e:EesA}), fulfills the following relation:$^{10)}$  
\begin{equation}\label{e:dEesdlam}
E_{\rm es}
-\sum {\bf R}_{\alpha}\!\cdot\!{\bf F}_\alpha
=-\int_V n({\bf r}){\bf r}\cdot\nabla{\delta {E}_{\rm es}\over\delta n({\bf r})}d{\bf r}\,,
\end{equation}
with the force  ${\bf F}_\alpha\equiv -\nabla_\alpha E_{\rm es}$ on the $\alpha$th nucleus, 
and the effective external potential is defined by 
$U_{\rm eff}({\bf r})
={\delta [E_{\rm es}+{E}_{\rm xc}]}/{\delta n({\bf r})}$,
Janak has derived from (\ref{e:TsTr2}) and (\ref{e:dExcdlam2})
\begin{eqnarray}
\oint_{\partial V}{\bf r}\cdot({\bf\textsf P}_{\rm K}^{\rm DF}+{\bf\textsf P}_{\rm xc})\cdot d{\bf S}+\sum {\bf R}_{\alpha}\!\cdot\!{\bf F}_\alpha&=&2T_s[n]+E_{\rm es}+\int_V {\rm tr}{\bf\textsf P}_{\rm xc}d{\bf r} \label{e:Ja} \\
&=&2T_{\rm e}+U_{\rm e}\,,
\end{eqnarray}
with {\it neglect} of the surface-integral term in (\ref{e:Ja}) as mentioned before. 
If we put $3PV\equiv \sum {\bf R}_{\alpha}\!\cdot\!{\bf F}_\alpha$ in (\ref{e:Ja}) as Janak defined, we obtain
\begin{eqnarray}
3PV=2T_s[n]+E_{\rm es}+\int_V {\rm tr}{\bf\textsf P}_{\rm xc}d{\bf r}-\oint_{\partial V}{\bf r}\cdot({\bf\textsf P}_{\rm K}^{\rm DF}+{\bf\textsf P}_{\rm xc})\cdot d{\bf S} \label{e:JanP} \,.
\end{eqnarray}
Therefore, {\it this pressure definition provides neither the electron pressure nor the total pressure of System-I in the Janak derivation}.
On the other hand, if we put in the above
\begin{equation}
3\tilde{P}_{\rm e}V\equiv \oint_{\partial V}{\bf r}\cdot({\bf\textsf P}_{\rm K}^{\rm DF}+{\bf\textsf P}_{\rm xc})\cdot d{\bf S}=2T_{\rm e}+U_{\rm e}-\sum {\bf R}_{\alpha}\!\cdot\!{\bf F}_\alpha=0\,,
\end{equation}
the pressure given by (\ref{e:JanP}) becomes the total pressure of System-II with infinite nuclear mass.
In contrast with the above fact,  Eq.~(\ref{e:Ja}) {\it with} the surface-integral term is identical with (\ref{e:PeI}) or (\ref{e:Om}): 
the sum of the electron and nuclear pressures are equal to 
the total pressure of System-I with infinite nuclear mass. 
In reality, Janak's pressure formula for numerical calculation becomes that of System-I, since the surface-integral terms in (\ref{e:TsTr2}) and (\ref{e:dExcdlam2}), in comparison with his equation~(7), are taken into account in his equation (23) at this stage.


\begin{thebibliography}{99}
\bibitem{Janak74} J. F. Janak,  
\PRB{9,1974,3985-3988}; J. F. Janak, V. L. Moruzzi, and A. R. Williams, \PRB{12,1975,1257-1261}
\bibitem{Averill81} F. W. Averill and G. S. Painter,  
\PRB{24,1981,6795-6800}.
\bibitem{SlaterBK2} J. C. Slater, 
 {\it Quantum theory of molecules and solids} 
vol.4 (McGraw-Hill, 1974) p.~291.
\bibitem{ZiescheGN88} P. Ziesche, J. Grafenstein, and O. H. Nielsen,  
\PRB{37,1988,8167-8178}.
\bibitem{Price71} D. L. Price,  
\PRA{4,1971,358-363}.
\bibitem{Hafner87} J. Hafner, 
{\it From Hamiltonians to Phase Diagrams} (Springer, 1987).
\bibitem{Perdew90} J. P. Perdew, H. Q. Tran, and E. D. Smith,  
\PRB{42,1990,11627-11636}.
\bibitem{Shore91} H. B. Shore and J. H. Rose,  
\PRL{66,1991,2519-2522}.
\bibitem{Shore99} H. B. Shore and J. H. Rose,  
\PRB{59,1999,10485-10492}.
\bibitem{ChiharaUn} J. Chihara and M. Yamagiwa, 
\PTP{111,2004,339-359}.
\bibitem{SlaterBK1} J. C. Slater, {\it Quantum theory of molecules and solids}
vol.3 (McGraw-Hill, 1967) p.~349.
\bibitem{Mason69} E. A. Mason and T. H. Spurling, {\it The virial equation of state} (Pergamon, 1969).
\bibitem{Zwanzig50} R. W. Zwanzig,  
{J. Chem. Phys.} {\textbf 18}  (1950), 1412-1413.
\bibitem{Brown58} W. B. Brown,  
{J. Chem. Phys.} {\textbf 28} (1957), 522-523.
\bibitem{Marc85} G. Marc and W. G. McMillan, 
 {\it Advances in Chemical Physics} vol. 58  (1985), 209-361.
\bibitem{RiddelUlen50} R. J. Riddel, Jr. and G. E. Uhlenbeck, 
{J. Chem. Phys.} {\textbf 18} (1950), 1066-1069. 
\bibitem{Nishiyama51} T. Nishiyama,  
{J. Chem. Phys.} {\textbf 19}  (1951), 1320-1320.
\bibitem{NM85} O. H. Nielsen and R. M. Martin,  
\PRB{32,1985,3780-3791}.
\bibitem{Corso94} A. Dal Corso and R. Resta,  
\PRB{50,1994,4327-4331}
\bibitem{Chihara} J. Chihara, {J. of Phys.:Condens. Matter} {\textbf 3} (1991), 8715-8744.
\bibitem{Bader80} R. F. W. Bader  
{J. Chem. Phys.} {\textbf 73}  (1980), 2871-2883.     
\bibitem{SrebBader75} S. Srebrenik and R. F. W. Bader,  
{J. Chem. Phys.} {\textbf 63} (1975),  3945-3961.  
\bibitem{BaderAus} R. F. W. Bader and M. A. Austen,  
{J. Chem. Phys.} {\textbf 107}  (1997), 4271-4285.   
\bibitem{More79} R. M. More,  
\PRA{19,1979,1234-1246}.
\bibitem{More85} R. M. More,  
{\it Adv. Atom. and Mol. Phys.} vol. {\bf 21}  (1985), 305-356.    
\bibitem{Barto80} L. J. Bartolotti and R. G. Parr,  
{J. Chem. Phys.} {\textbf 72} (1980), 1593-1596.    
\bibitem{Chihara01} J. Chihara, I. Fukumoto, M. Yamagiwa, and H. Totsuji, 
{ J. of Phys.:Condens. Matter} {\bf 13} (2001), 7183-7198.
\bibitem{Folland86} N. O. Folland, 
\PRB{34,1986,8305-8312}.
\bibitem{MackAnd75} A. R. Mackintosh and O. K. Andersen,  
in {\it Electrons at the Fermi surfaces}, ed. M. Springford 
(Cambridge University Press, 1975).
\bibitem{Skriver84} H. L. Skriver, 
{\it The LMTO method} (Springer-Verlag, 1984).
\bibitem{Liberman71} D. A. Liberman, \PRB{32,1971,2081-2082}.
\bibitem{Pettifor76} D. G. Pettifor, {\it Commun. on Phys.} {\bf 1}, (1976), 
141-146. 
\bibitem{LegPerrot01} P. Legrand and F. Perrot,   
{ J. of Phys.; Condens. Matter} {\bf 13}  (2001), 287-301.
\bibitem{R69} W. E. Rudge, 
\PR{181,1969,1033-1035}.
\bibitem{W79} A. R. Williams, J. Kuebler, and C. D. Gelatt,  
\PRB{19,1979,6094-6118}.
\bibitem{Hansen86} J. P. Hansen and I. R. McDonald,  
{\it Theory of simple liquids} (Academic Press, 1986).
\bibitem{Chihara89} J. Chihara,  
\PRA{40,1989,4507-4516}.
\bibitem{NiemHodg76} R. M. Nieminen and C. H. Hodges,  
{J. of Phys.} F{\textbf 6}  (1976), 573-585.
\bibitem{Meyer71} A. Meyer, I. H. Umar, and W. H. Young, 
\PRB{4,1971,3287-3291}.
\bibitem{P79} F. Perrot,  
\PRA{20,1979,586-594}.
\bibitem{Heine72} V. Heine and C. H. Hodges,  
{J. of Phys.} C{\textbf 5}  (1972), 225-230.
\bibitem{MWYZ88} R. M. More, K. H. Warren, D. A. Young, and G. B. Zimmerman,  
{Phys. Fluids} {\textbf 31}  (1988), 3059-3078.
\bibitem{Kamb96} S. Kambayashi and J. Chihara,  
\PRE{53,1996,6253-6263}.
\bibitem{Kuma83} R. Kumaravadivel,  
{J. of Phys.} F{\bf 13}  (1983), 1607-1626.   
\bibitem{Copley77} J. R. D. Copley and S. W. Lovesey,  in 
{\it Liquid metal, 1976} [IOP Conf. Proc. No. 30], ed. R. Evans and D. A. Greenwood 
(The Institute of Physics, 1977), p.~171.
\bibitem{Waseda80} Y. Waseda, 
{\it The structure of non-crystalline materials} (McGraw-Hill, 1980).
\end{thebibliography}
\end{document}